\documentclass[letterpaper,11pt,leqno]{article}
\usepackage{paper, math}
\hypersetup{pdftitle={Workers as Partners: a Theory of
Responsible Firms in Labor Markets}}

\begin{document}

% Enter title:
\title{Workers as Partners: a Theory of Responsible Firms in Labor Markets}

% Enter authors:
\author{Francesco Del Prato, Marc Fleurbaey
%
% Enter affiliations and acknowledgements:
\thanks{Francesco Del Prato: \emph{Department of Economics and Business Economics, Aarhus University} (\href{mailto:francesco.delprato@econ.au.dk}{francesco.delprato@econ.au.dk}). 
Marc Fleurbaey: \emph{Paris School of Economics} (\href{mailto:marc.fleurbaey@psemail.eu}{marc.fleurbaey@psemail.eu}).
We thank Edoardo M. Acabbi, Axel Gottfries and the participants of the seminar of the \emph{“Measuring the Economy”} research chair for the useful comments.
The authors acknowledge the PSE research chair \emph{“Measuring the Economy”} for the financial support.
}}

% Enter date:
\date{February 2026} 

% Enter permanent URL (can be commented out):
%\available{https://francescodelprato.github.io/research/RF_theory.pdf}

\begin{titlepage}
\maketitle

% Enter abstract:
What happens when employers value worker welfare in frictional labor markets? We show this “responsibility” creates an endogenous wedge in the marginal labor cost—akin to a hiring subsidy—altering wage and vacancy incentives rather than only changing the surplus split. The wedge is strongest when outside options are weak and separations rare, implying larger wage premia in slack, low-mobility markets. In a wage-posting model with on-the-job search, responsible firms may occupy the high-wage segment even when less productive. In a DMP model, responsible firms commit to higher worker bargaining power, raising the value of unemployment and thereby wages at regular firms.

\vspace{1cm}
{\footnotesize{\color{gray}\textsc{Keywords}}: corporate social responsibility, search model, labor market, monopsony power.}

{\footnotesize{\color{gray}\textsc{JEL Classification}}: J2, J3, J5, J6, M14.}

\end{titlepage}

% Enter main text:
\section{Introduction}

What if firms do not treat wages as a pure cost, but also value the well-being of the people they employ? 
Recent stakeholder theory argues that when a firm puts weight on stakeholder value alongside profits, market power should be used less aggressively—indeed, in standard product-market settings, a stakeholder-oriented (“responsible”) firm may behave as if it were in perfect competition, using the competitive price as a reference point rather than extracting a markup \citep{magill2015theory, fleurbaey2023stakeholder}. 
Labor markets, however, pose a distinct challenge. 
Search frictions give employers wage-setting power even when product markets are competitive, and there is no external “competitive wage” that a firm can simply adopt as a responsible benchmark.

In frictional labor markets, wages are equilibrium objects shaped by firms’ wage and vacancy policies, workers’ outside options, and turnover. 
Worker welfare is inherently dynamic: the value of a job depends not only on today’s wage but also on future job-finding prospects, separation risk, and on-the-job mobility. 
As a result, “responsible wage setting” cannot be reduced to a static rule such as paying above-market wages, nor can it be defined by reference to a competitive price that exists outside the model. 
This raises a distinct set of questions: how should a responsible employer set wages when frictions generate monopsony power, and how should this vary with outside options and turnover?
What happens in equilibrium when responsible and profit-maximizing firms coexist?

We formalize responsibility as partial internalization of incumbent-worker surplus in canonical frictional labor-market models.
A responsible firm (REF) values profits and the per-period welfare of its employed workforce; the strength of this motive is governed by a single parameter.
This definition is deliberately narrow: it captures the idea of "workers as partners" in an ongoing employment relationship and focuses attention on the key frictional wedge between the wage bill and the worker surplus generated by a match.
It also separates what is internalized (incumbent-worker welfare) from how responsibility maps into policy levers: when wages are posted, responsibility directly affects wage and vacancy choices; when wages are bargained, it takes the form of committing to a more worker-friendly protocol.
We study this objective in three complementary environments: a firm-level benchmark with frictional hiring and efficiency wages, a wage-posting model with on-the-job search, and a Diamond--Mortensen--Pissarides (DMP) model with bargaining and endogenous tightness.

\paragraph{Responsibility as an endogenous labor-cost wedge}
A central mechanism of this paper is that responsibility in frictional labor markets is not only distributional. 
When a firm internalizes part of the surplus that employment generates for its incumbent workforce, raising wages increases the welfare term it values, and expanding employment increases the number of workers whose surplus it internalizes. 
In the firm’s optimality conditions, this appears as an endogenous wedge that lowers the effective marginal cost of labor—akin to a shadow subsidy on hiring. 
Put differently, a responsible firm behaves as if it faced a more elastic labor supply—and, in a precise limit, as if it were hiring at an effective wage below the posted wage.
This “shadow-subsidy” channel implies that doing good is not merely transferring rents from shareholders to workers; it also changes the firm’s perceived marginal incentives and can tilt wage and vacancy choices toward expansion.
Importantly, the magnitude of this wedge is state-dependent. 
When outside options are weak and jobs are persistent, the welfare gain from a higher wage is large and long-lived, so the responsibility wedge is strongest. 
When outside options are strong or turnover is high, the marginal welfare gain from wage increases is smaller, and the responsible firm’s behavior converges toward that of a standard profit-maximizing monopsonist. 
This delivers a sharp comparative-statics prediction: responsible wage premia should be larger precisely in environments where workers are more “captive,” i.e., when mobility is low and outside options are weak.

Stakeholder objectives in the labor market also raise an immediate concern. 
The appeal of profit maximization is its simplicity and observability: it provides a clear metric against which decisions can be evaluated. 
Once firms trade off profits against worker welfare, concerns arise about arbitrariness, agency problems, and the risk that almost any action can be rationalized post hoc \citep{mayer2013firm, de2020elephant}.
These concerns are amplified in labor markets because welfare is dynamic and depends on future opportunities. 
A goal of this paper is therefore to provide a tractable and testable theory of responsible wage setting in frictional labor markets—one that pins down how a well-specified responsibility motive enters canonical labor-market environments and yields comparative-statics predictions that are directly tied to measurable objects (outside options, turnover, mobility).

\paragraph{Main results} 

The paper delivers three sets of results.

First, in the firm-level benchmark, responsible firms' distinctive behavior---higher wages and stronger hiring incentives relative to profit-maximizers---is most pronounced in slack labor markets, where the responsibility wedge is largest, suggesting a countercyclical dimension to responsible wage setting.
These predictions are consistent with evidence that CSR initiatives can affect turnover and productivity \citep{flammer2015does} and that employee-friendly firms can outperform their peers \citep{edmans2011does}.\footnote{%
The empirical evidence on CSR and worker outcomes is suggestive rather than definitive; our focus is on providing a tractable mechanism within canonical search-and-matching models.}
A particularly transparent implication arises in the limiting case where labor supply becomes very elastic: profit-maximizing firms treat the wage as an exogenous market rate, while responsible firms behave as if they faced a lower effective wage benchmark that nets out the surplus they internalize---showing that responsibility can reshape behavior even when wage-setting power becomes negligible in the usual elasticity sense.

Second, responsibility leaves a clear footprint on the wage distribution and the segmentation of firm types. In the wage-posting model with on-the-job search, responsible firms are pulled toward higher wages because higher wages both attract and retain workers and raise the welfare term the firm internalizes. This force can generate a segmented high-wage sector populated by responsible firms, even when responsibility is partial and even when responsible firms are less productive than their profit-maximizing counterparts, depending on productivity differences and the strength of the efficiency-wage channel. In the limit of full responsibility, responsible firms’ wage offers bunch: all responsible firms optimally choose the same wage, either the wage that maximizes total match surplus or the highest sustainable wage implied by solvency (wages cannot exceed output). Partial responsibility thus generates wage dispersion and stratification across segments, while full responsibility tends toward wage compression within the responsible-firm segment.

Third, responsibility interacts with market tightness, entry, and wage spillovers. 
In the DMP setting, there is no direct “wage policy” inside a match beyond the bargaining protocol, so we model responsibility as a commitment to higher worker bargaining power. 
Normatively, shifting bargaining power toward workers can raise per-match stakeholder value in slack markets, because workers’ continuation prospects are weaker than firms’ vacancy-filling prospects; but under free entry this comes at a cost: higher wages reduce profitability, equilibrium tightness, and the number of operating firms, creating a sharp employment trade-off. 
Positively, responsible firms that commit to a more worker-friendly bargaining protocol raise the value of unemployment and push up wages at ordinary firms through improved outside options. 
As the fraction of responsible firms increases, the wage gap shrinks because ordinary-firm wages respond more strongly. 
Under the maintained assumptions of the DMP model (common vacancy costs and filling probabilities), responsible firms cannot coexist with profit-maximizers under free entry; we therefore focus on limited competition.\footnote{%
Endogenizing the retention and recruiting advantages identified in Section~\ref{sec:micro_model} could sustain coexistence---a direction we leave for future work.}

\paragraph{Contribution to the literature}
We contribute to three strands of literature.
First, we provide a theory of CSR in frictional labor markets within standard search-and-matching frameworks.
Recent work incorporates CSR into quantitative general equilibrium models \citep{alfaro2022responsible}, industrial organization \citep{allcott2023economic}, and stakeholder-firm theory \citep{magill2015theory, hart2017companies, fleurbaey2023stakeholder}.
Relative to stakeholder-firm theories that consider broad sets of stakeholders, we focus on workers and embed responsibility into canonical labor-market environments with monopsony power, studying how responsibility interacts with frictions, wage posting, effort incentives, and bargaining to shape wages, employment, and market power.
In this sense, we complement empirical evidence on CSR and worker outcomes \citep{boudreau2024multinational, hedblom2019toward, list2021corporate} by providing mechanisms in a familiar and tractable environment.

Second, building on work on CSR and competition \citep[e.g.,][]{planer2020strategic}, we study coexistence and survival in frictional labor markets.
While some contributions emphasize selection mechanisms that discipline or eliminate non-responsible firms \citep{brekke2008attracting}, our analysis highlights how REF wage-setting affects segmentation and coexistence, and how these outcomes differ under free entry versus limited competition.

Third, we analyze market-wide effects of responsible wage setting.
Rather than relying on external support or specific organizational forms \citep{baron2008managerial, fioretti2022caring, besley2017profit}, we show how higher wages at responsible firms can raise the value of unemployment and thereby increase wages economy-wide through improved outside options.
Under limited competition, this channel allows REFs to coexist with ordinary firms while enhancing worker welfare through standard market mechanisms.

\paragraph{Roadmap}
Section~\ref{sec:micro_model} presents the firm-level benchmark.
Section~\ref{sec:BM} studies wage dispersion and segmentation in a wage-posting model.
Section~\ref{sec:DMP} studies corporate governance and equilibrium employment in a DMP model.
Section~\ref{sec:conclusions} concludes.
Formal proofs are in the Appendix.

\section{Labor market responsibility at the firm level}\label{sec:micro_model}
Here, we introduce a model of firm behavior that features corporate social responsibility in the presence of labor market frictions. 
The aim is to provide a tractable benchmark for how responsibility affects wage and vacancy choices when hiring is frictional and wages influence productivity through efficiency wages.

\subsection{Environment and setting}
Consider a firm operating across multiple labor market segments, each characterized by workers with different qualifications. 
Time is discrete, and the analysis focuses on the steady state equilibrium.
The firm's production function depends on both the quantity of workers employed and their wage-dependent productivity: $y = f \of{\ell, w}$, where $\ell = \bp{\ell_1, \dots, \ell_m}$ represents the vector of workers across $m$ qualification categories, and $w = \bp{w_1, \dots, w_m}$ denotes the corresponding wages. We assume that working hours are fixed exogenously.

Worker productivity is influenced by wages through an efficiency wage effect:
\[ 
    y = f \of{\ell, w} = f \of{\ell_1 e_1 \of{w_1}, \dots, \ell_m e_m \of{w_m}}
\]
The effort functions $e_k(w_k)$ for $k = 1, \dots, m$ are increasing in wages, capturing the motivational effect of higher compensation on worker productivity.
\footnote{
We provide a formal proof of this statement in Lemma \ref{lemma:e_in_w}.
}

To recruit workers, the firm must post vacancies $v = \bp{v_1, \dots, v_m}$ and offer corresponding wages $w$. The market-level tightness is equal to the ratio of vacancies $\bar{v}_k$ (aggregating vacancies over all firms) over unemployment $u_k$ in segment $k$: $\theta_k = \bar{v}_k/u_k$.
	We summarize hiring in segment $k$ by the expected number of filled jobs:
	\[
	\E{\ell_k}=v_k\,\bar q_k(\theta_k,w_k).
	\]
	We write $\E{\ell_k}$ for expected employment in segment $k$ and suppress its dependence on $(v_k,w_k)$ when no ambiguity arises.
	The per-vacancy steady-state fill probability satisfies $\bar q_k(\theta_k,w_k)\in(0,1)$ and is decreasing in tightness:\footnote{%
	In standard matching environments, unemployed workers' job finding typically moves in the opposite direction: if $\hat p_k(\theta_k,w_k)=\theta_k\,\hat q_k(\theta_k,w_k)$, then $\pd{\hat p_k(\theta_k,w_k)}{\theta_k}>0$ even though $\pd{\bar q_k(\theta_k,w_k)}{\theta_k}<0$.}
\[
\pd{\bar q_k(\theta_k,w_k)}{\theta_k}<0.
\]
We allow it to depend on $w_k$, reflecting a greater ease to attract workers with higher wages, for a given level of market tightness.

We define $\bar q_k$ from primitives as follows. Let $\hat q_k(\theta_k,w_k)\in(0,1)$ denote the per-period meeting probability for an open vacancy (decreasing in $\theta_k$) and let $\gamma_k(w_k)\in(0,1)$ denote the per-period separation probability for an occupied job.
The per-vacancy steady-state fill probability is:
\[
\bar q_k(\theta_k,w_k):=\frac{\hat q_k(\theta_k,w_k)}{\hat q_k(\theta_k,w_k)+\gamma_k(w_k)}.
\]
This expression follows from steady-state flow balance between new matches and separations.

\paragraph{Timing}
Wages and vacancies are posted $(w_k,v_k)$. 
Then, matching occurs and employed workers choose effort $e_k(w_k)$. 
Production and payoffs occur, and separations realize at the end of the period.
Throughout, we write the separation probability as $\gamma_k(w_k):=\gamma_k(e_k(w_k))$.

% Worker turnover in segment $k$ occurs at a rate $\d_k\of{w_k, \gamma_k(e_k(w_k))} \equiv \d_k (w_k)$, a fixed cost $\k$, where $\gamma_k(\cdot)$ is the separation probability perceived by the worker.
% The proportion $\d_k(w_k)$ decreases with both wage $w_k$ and effort $e_k$, reflecting that higher wages improve retention and higher effort reduces dismissal likelihood. 
% The turnover function is treated as exogenous, meaning the firm cannot implement different dismissal policies.

\subsection{Profit-maximizing firm}
The expected profit of a firm posting vacancies $v$ and offering wages $w$ is:\footnote{
For simplicity, we ignore the specific cost of posting a vacancy in this section, as it does not induce any difference between the behavior of profit-maximizing firms and REFs.
}
\begin{equation}\label{eq:profit}
	    P\of{v,w} = \E{f\of{\ell,w}} - \sum_{k=1}^m w_k\,\E{\ell_k}
\end{equation}
The first term is expected production, taken over the distribution of filled jobs $\ell$, and the second term is the expected wage bill across segments.\footnote{%
Appendix \ref{a:add_computations_micro} provides an explicit representation of the expectation and derives the first-order conditions.}
The firm chooses $(v,w)$ to maximize $P(v,w)$.

\paragraph{Optimal vacancy posting} 
The first order condition for vacancy posting in market segment $k$ is:
\begin{equation}\label{eq:FOC_v}
	    \pd{}{v_k}\E{f\of{\ell,w}} = w_k \pd{\E{\ell_k}}{v_k}.
\end{equation}
noting that $\E{\ell_k} = v_k\,\bar q_k(\theta_k,w_k)$ with $\theta_k=\bar{v}_k/u_k$.
Taking $u_k$ as given, the tightness channel captures a congestion effect: posting more vacancies raises tightness $\theta_k$, which lowers the per-vacancy fill probability $\bar q_k(\theta_k,w_k)$.
As a result, the marginal employment gain from an additional vacancy is smaller than $\bar q_k$ and given by
\[
    \pd{\E{\ell_k}}{v_k}
    = \bar q_k(\theta_k,w_k) + v_k\,\pd{\bar q_k(\theta_k,w_k)}{\theta_k}\pd{\theta_k}{v_k}
    = \bar q_k(\theta_k,w_k) + \theta_k\,\pd{\bar q_k(\theta_k,w_k)}{\theta_k},
\]
taking account of the fact that $\partial \bar{v}_k/\partial v_k = 1$.
We impose a mild regularity condition ensuring that this congestion is not so strong as to make additional vacancies reduce expected employment:
\[
    \bar q_k(\theta_k,w_k) + \theta_k\,\pd{\bar q_k(\theta_k,w_k)}{\theta_k} > 0
    \quad\iff\quad
    \varepsilon_{\bar q_k:\theta_k}(\theta_k,w_k)<1,
\]
where 
$\varepsilon_{\bar q_k:\theta_k}(\theta_k,w_k):= -\frac{\theta_k}{\bar q_k(\theta_k,w_k)}\pd{\bar q_k(\theta_k,w_k)}{\theta_k}$
is the elasticity of the per‑vacancy fill probability with respect to tightness $\theta_k$.\footnote{%
For example, if $\hat q_k(\theta_k,w_k)=A_k(w_k)\theta_k^{-\alpha_k}$ with $\alpha_k\in(0,1)$ and $\gamma_k$ does not depend on $\theta_k$, then $\varepsilon_{\bar q_k:\theta_k}<\alpha_k<1$.}

Condition \eqref{eq:FOC_v} balances the marginal contribution of an additional vacancy to expected output with the marginal wage bill induced by the resulting increase in expected employment.

\paragraph{Optimal wage posting}
The first order condition for wage setting in market segment $k$ is 
\begin{equation}\label{eq:FOC_w}
    \pd{}{w_k}\E{f\of{\ell,w}}=\E{\ell_k}+w_k\,\pd{\E{\ell_k}}{w_k}.
\end{equation}
This condition balances the marginal benefits and costs of wage increases. 
Increasing $w_k$ can raise expected output (through improved hiring/retention and higher effort) but it also raises the expected wage bill both mechanically (through the existing workforce) and through the induced change in employment.
% The right-hand side of the condition \eqref{eq:FOC_w} can be reformulated as
% \begin{equation}\label{eq:FOC_w_rhs}
%     w_k \pd{}{w_k} \E{\ell_k\of{v_k, w_k}} + \E{\ell_k\of{v_k, w_k}} 
% \end{equation}

Define the elasticity of expected labor supply to wages as:
\[
    \varepsilon_{\E \ell_k : w_k} \equiv \frac{w_k}{\E{\ell_k}} \pd{\E{\ell_k}}{w_k}
\]
Using this elasticity, the first-order condition \eqref{eq:FOC_w} can be rewritten as:
\begin{equation}\label{eq:FOC_w_elast}
\frac{\pd{}{w_k}\E{f\of{\ell,w}}}{\pd{\E{\ell_k}}{w_k}}=
\bp{1+\frac{1}{\varepsilon_{\E \ell_k:w_k}}} w_k
\end{equation}
This expression shows how the optimal wage depends on the elasticity of expected labor supply. 
When $\varepsilon_{\E \ell_k : w_k}$ is high (labor supply is elastic), the markup over the marginal product of labor approaches zero, resembling a competitive outcome. 
Conversely, when this elasticity is low (labor supply is inelastic), the firm exerts greater monopsony power, setting wages below the marginal revenue product by a larger margin. 

% This expression rewrites the marginal cost in terms of labor supply elasticity.
% A higher $\varepsilon_{\E \ell_k : w_k}$ implies that hiring responds more strongly to wage changes, reducing the per-unit cost of wage adjustments.

We next introduce the worker values and surplus, which will enter the responsible firm's objective.

\subsection{Worker surplus}
The worker's decision-making process in the labor market is captured by a discounted present value that accounts for current compensation, effort costs, and future employment prospects.

\paragraph{Employment}
The present value of employment for a worker in segment $k$ receiving wage $w_k$ is
\begin{equation}\label{eq:worker_surplus}
    \begin{split}
        W_k(w_k) 
        &= w_k - c_k\of{e_k(w_k)}
        + \beta  \Bigl[
        \gamma_k(w_k) U_k  \\
        &+ \bp{1 - \gamma_k(w_k)}
        \bs{p_k(w_k) \E[w > w_k]{W_k(w)}
        + \bp{1 - p_k(w_k)} W_k(w_k) }
        \Bigr]
    \end{split}
\end{equation}
where $c_k\of{e_k(w_k)}$ represents the cost of effort, $p_k(w_k)$ is the probability of receiving an improving job offer, $\beta < 1$ is the discount factor, and $\E[w > w_k]{W_k(w)}$ represents the expected value of getting a wage above $w_k$ in segment $k$.

The worker's present value \eqref{eq:worker_surplus} consists of current net income plus a discounted factor that reflects three possible future states: unemployment, obtaining a higher-paying job, or remaining in the current position.
We allow offer arrival rates to differ between unemployment and employment.
Let $\bar{p}_k^{u}$ denote the probability that an unemployed worker in segment $k$ receives an offer in a period, and let $\bar{p}_k^{e}$ denote the corresponding on-the-job offer arrival probability.
Offers are drawn from the segment-specific distribution $F_k(\cdot)$.
Accordingly, the probability that an employed worker with wage $w_k$ receives an offer exceeding $w_k$ is
$p_k(w_k)=\bp{1 - F_k(w_k)} \bar{p}_k^{e}$.

\paragraph{Unemployment}
When unemployed, the worker's present value in segment $k$ is:
\[
U_k = b + \beta \bs{ (1 - \bar{p}_k^{u}) U_k + \bar{p}_k^{u} \E{W_k(w)} } 
= \frac{b + \beta \bar{p}_k^{u} \E{W_k(w)}}{1 - \beta (1 - \bar{p}_k^{u})}
\]
This value function includes the per-period unemployment benefit $b$ and a continuation value reflecting the probability of receiving a job offer in segment $k$ and drawing a wage from $F_k$.

\paragraph{Surplus}
Worker surplus (per employed worker) is defined as the present-value gain from employment relative to unemployment:
\begin{equation}\label{eq:surplus}
S_k(w_k):=W_k(w_k)-U_k.
\end{equation}
The worker chooses effort to maximize $W_k(w_k)$ (equivalently $S_k(w_k)$), taking $U_k$ and the continuation term $\E[w > w_k]{W_k(w)}$ as given.
When $S_k$ enters the firm's objective (Section~\ref{ss:ref_micro}), it is scaled by $(1-\beta)$ to convert the present value into a per-period flow.

We will repeatedly use two monotonicity properties of the worker problem: wages increase the value of employment, and (under our maintained curvature assumptions) higher wages induce higher effort. The proofs of the following results are in Appendix~\ref{a:add_computations_micro}.

\begin{lemma}\label{lemma:W_in_w}
    The function $W_k(w)$ is increasing in $w$.
\end{lemma}

\begin{lemma}\label{lemma:e_in_w}
    The function $e_k(w)$ is increasing in $w$, assuming that $c_k$ is convex in effort and that $\gamma_k$ is decreasing and convex in effort.
\end{lemma}
These monotonicities imply that worker surplus responds to wages, outside options, and turnover in intuitive ways:

\begin{proposition}\label{prop:surplus_behavior}
    Holding $\bar p_k^{e}$ fixed, the surplus $S_k(w_k)$ is increasing in $w_k$, decreasing in the separation probability $\gamma_k(w_k)$, and decreasing in the unemployment offer arrival probability $\bar{p}_k^{u}$.
\end{proposition}

Proposition \ref{prop:surplus_behavior} formalizes two forces that will matter throughout the paper.
A higher unemployment offer arrival rate $\bar p_k^{u}$ raises the unemployment value $U_k$ and therefore compresses the gain from employment, while a higher separation probability $\gamma_k(w_k)$ lowers the continuation value of a job.
As we show below, because responsible firms internalize worker surplus, stronger outside options and higher turnover will mechanically weaken the responsibility motive and shrink the responsibility wedge.

\subsection{The responsible firm}\label{ss:ref_micro}
A responsible firm (REF) values the present value of the expected profit flow and the welfare of its incumbent workers, measured by their surplus:\footnote{%
This reflects a departure from pure profit maximization, aligned with non-strategic CSR concepts discussed in \cite{kitzmueller2012economic}.}
\[
    \max_{(v,w)}\ \frac{1}{1-\beta}P(v,w)+\eta\sum_{k=1}^m S_k(w_k)\,\E{\ell_k}.
\]
The parameter $\eta\ge 0$ governs the strength of responsibility ($\eta=0$ recovers the standard profit-maximizing firm).

In segment $k$, the REF values incumbent-worker welfare each period, proportional to the size of its workforce $\E{\ell_k}$; responsibility is therefore a stock objective tied to ongoing employment relationships.
This also means the responsibility term creates a hiring incentive: under our maintained regularity condition, posting more vacancies raises expected employment $\E{\ell_k}$, increasing the number of workers whose surplus the firm values.\footnote{%
An alternative would be to weight only marginal hires; we focus on incumbent-worker welfare because it maps naturally into the firm’s wage bill and the within-match wedge.}
We assume homogeneity within each $k$, with identical surplus $W_k$ and effort level $e_k$, which is incorporated in the production function $f$.

As a helpful benchmark, we briefly recall how a REF behaves in a \emph{frictionless competitive benchmark}.

\paragraph{Frictionless competitive benchmark}
In this benchmark, the weighted surplus can be expressed as:
\[
    f\of{\ell(w)} - w \ell(w) + \eta\sum_k \mathcal{S}_k(w_k)
\]
where $f\of{\ell(w)}$ represents the production value and $w \ell(w)$ represents the total wage bill. Here, $\mathcal{S}_k(w_k)$ denotes workers' \emph{total} surplus in segment $k$:
\[
    \mathcal{S}_k(w_k) = w_k \ell_k(w_k) - \int_0^{\ell_k(w_k)} \ell^{-1}_k(l) dl,
\]
where $\ell^{-1}_k(l)$ is the inverse labor supply function representing the reservation wage in market segment $k$. 
The integral term calculates the total reservation wage cost for all employed workers.

The objective of the firm then reads as:
\[
    f\of{\ell(w)} - (1-\eta) \sum_k w_k \ell_k(w_k) - \eta \sum_k \int_0^{\ell_k(w_k)} \ell^{-1}_k(l) dl
\]
This entails the following first-order condition for wage setting:
\[
    f_k\of{\ell(w)} = w_k \left( 1 + (1-\eta) \frac{1}{\varepsilon_{\ell_k : w_k}} \right)
\]
which, in the "fully responsible" case $\eta=1$, simplifies to the perfectly competitive condition equating the marginal product of labor with the wage. In the general case, the parameter $\eta>0$ reduces the use of market power by dampening the elasticity term in the equation.

We now examine how this standard result is modified in the presence of the frictions introduced by our current model.
Detailed derivations are provided in Appendix \ref{a:add_computations_micro}.

\paragraph{Optimal REF vacancy posting in the current model}
The first-order condition for optimal vacancy posting for the responsible firm is
\begin{equation}\label{eq:FOC_v_REF}
    \pd{}{v_k}\E{f\of{\ell,w}}
    = \bs{w_k - \eta(1-\beta) S_k(w_k)} \pd{\E{\ell_k}}{v_k}.
\end{equation}
Relative to the profit-maximizing firm's condition \eqref{eq:FOC_v}, this expression replaces the wage $w_k$ by the effective wage cost $w_k-\eta(1-\beta)S_k(w_k)$, i.e., it embeds the per-period worker-surplus flow $(1-\beta)S_k(w_k)$ in the firm's marginal cost of expanding employment.
This yields:

\begin{proposition}\label{prop:REF_vacancies}
    The responsible firm posts vacancies analogously to a profit-maximizing firm but with a wage rate reduced by $\eta(1-\beta) S_k(w_k)$.
\end{proposition}

The responsible firm effectively operates with a reference wage below the market wage $w_k$, as it accounts for workers’ surplus $S_k$.
The magnitude of this reduction depends on $\eta$ and on parameters influencing the surplus, as analyzed in Proposition \ref{prop:surplus_behavior}. 
Specifically, $S_k$ increases when unemployment offers are scarce (low $\bar{p}_k^{u}$) and separation risk is low (low $\gamma_k(w_k)$), reflecting weak outside options and low turnover, or when wages are high.
Two limiting cases illustrate this effect:
\begin{itemize}
    \item When $\eta=1$, $\gamma_k(w_k) = 0$, and there are no wage offers either in unemployment or on the job ($\bar{p}_k^{u}=\bar{p}_k^{e}=0$), the expression for this adjusted wage simplifies to:
    \[ w_k - (1-\beta) S_k = c_k \of{e_k (w_k)} + b \]
    meaning that a fully responsible firm posts vacancies as if the wage was equal to the workers’ reservation value, implying maximal consideration of worker welfare when outside options and turnover are weak.
    \item When $\eta=1$ and outside options and turnover are very high ($\bar p_k^{u}\simeq 1$ and $\gamma_k(w_k)\simeq 1$), jobs are short-lived and workers can readily move.
    In this case the employment surplus $S_k(w_k)$ is small (and therefore the per-period flow $(1-\beta)S_k(w_k)$ is negligible), so the reference wage closely tracks the market wage $w_k$ and vacancy posting is close to the profit-maximizing benchmark.
\end{itemize}
For $\eta<1$, the vacancy-posting wedge $\eta(1-\beta)S_k$ is proportionally smaller, so the reference wage lies closer to the market wage.

Thus, the responsible firm posts vacancies as if facing a wage between the market wage and the reservation wage. When outside options and turnover are high (high $\bar p_k^{u}$ and $\gamma_k$), the reference wage approaches the market wage $w_k$ and the firm’s behavior resembles that of a profit-maximizing firm. 
Conversely, when outside options are weak and turnover is low, the reference wage shifts toward the reservation value, reflecting greater dependence of workers on the firm.

\paragraph{Optimal REF wage posting in the current model}
The first-order condition for optimal wage posting by the responsible firm is:
\begin{equation}\label{eq:FOC_w_REF}
    \begin{split}
        &\frac{\pd{}{w_k}\E{f\of{\ell,w}}}{\pd{\E{\ell_k}}{w_k}}
        =
        \bp{1 + \frac{1}{\varepsilon_{\E \ell_k:w_k}}} w_k
        - \eta(1 - \beta) S_k(w_k) \bp{1 + \frac{\varepsilon_{S_k:w_k}}{\varepsilon_{\E \ell_k:w_k}}}
    \end{split}
\end{equation}
where
\[ \varepsilon_{S_k:w_k} \equiv \pd{S_k}{w_k} \frac{w_k}{S_k} \]
denotes the elasticity of worker surplus with respect to the wage.

A marginal wage increase raises the firm's expected wage bill on all employed workers (a direct cost proportional to $\E{\ell_k}$) and, through induced hiring, by $w_k\,\pd{\E{\ell_k}}{w_k}$.
Under responsibility, the same wage increase also increases the welfare of incumbent workers: per employed worker, the per-period surplus flow rises by $\eta(1-\beta)\,\pd{S_k}{w_k}$, and the additional employment created by the wage increase raises the weight placed on worker surplus by $\eta(1-\beta)S_k\,\pd{\E{\ell_k}}{w_k}$.
Netting these welfare gains against the wage-bill effect yields the subtractive term in \eqref{eq:FOC_w_REF}; equivalently, relative to a profit-maximizing firm, the REF behaves as if the effective wage cost were reduced.
Efficiency-wage effects enter both governance regimes symmetrically through $\pd{}{w_k}\E{f\of{\ell,w}}$ and therefore do not generate an additional wedge.

Thus, the responsible firm’s wage-setting strategy differs from that of a profit-maximizing firm in its approach to market power. 
While a profit-maximizing firm suppresses wages based on the inverse of the labor supply elasticity, the responsible firm behaves as if the labor supply was more elastic, elevating wages by incorporating worker welfare. 
The following proposition summarizes this effect:

\begin{proposition}\label{prop:REF_markdown}
    The responsible firm sets wages analogously to a profit-maximizing firm, but facing a labor supply elasticity scaled by:
  \begin{equation}\label{eq:varXi}
    \varXi = \frac{1}{D_k},
    \qquad
    D_k:=1 - \eta(1 - \beta)
    \left(\frac{S_k(w_k)}{w_k}\varepsilon_{\E \ell_k:w_k} + \pd{S_k(w_k)}{w_k}\right),
  \end{equation}
  resulting in higher posted wages.
\end{proposition}
We assume $D_k>0$ so that $\varXi$ is well-defined, i.e.
\[
\eta(1-\beta)\left(\frac{S_k(w_k)}{w_k}\varepsilon_{\E \ell_k:w_k}+\pd{S_k(w_k)}{w_k}\right)<1.
\]
This is a mild regularity condition that rules out cases in which responsibility is so strong (or employment rents so large) that the elasticity-scaling term would blow up. Under $S_k(w_k)>0$ and $\pd{S_k(w_k)}{w_k}>0$, we have $D_k<1$ and hence $\varXi>1$.
Finally, when $\eta=0$ we have $D_k=1$ and thus $\varXi=1$: the REF's wage condition collapses to that of a profit-maximizing firm (and similarly for vacancy posting).
The next corollary provides a more transparent economic decomposition, separating the benchmark wage shift from the elasticity scaling effect:

\begin{corollary}\label{cor:REF_wage_decomposition}
    The responsible firm behaves analogously to a profit-maximizing firm that would post wages equal to $w_k - \eta(1-\beta) S_k(w_k)$ and would be facing a labor supply elasticity multiplied by:
    \[ \varGamma = \frac{1-\eta(1 - \beta)\frac{S_k(w_k)}{w_k}}{1-\eta(1 - \beta)\pd{S_k(w_k)}{w_k}}. \]
\end{corollary}
This alternative formulation suggests that the responsible firm treats $w_k-\eta(1 - \beta)S_k(w_k)$ as a wage benchmark, similar to its approach in vacancy posting, but modulates its market power based on the surplus sensitivity ($\frac{\partial S_k}{\partial w_k}$) and the surplus-to-wage ratio ($\frac{S_k}{w_k}$).
Typically, $\varGamma$ rises with $\varepsilon_{S_k:w_k}$, indicating larger wage adjustments when the surplus is highly responsive to wage changes.
Moreover, $\varGamma>1$ is equivalent to $\pd{S_k(w_k)}{w_k}>\frac{S_k(w_k)}{w_k}$ (i.e.\ $\varepsilon_{S_k:w_k}>1$), which we treat as a maintained condition when interpreting $\varGamma$ as an elasticity scaling factor.

\paragraph{Market conditions and the REF's use of market power}
We now discuss how labor market conditions and wage levels influence the responsible firm’s exercise of market power.

The next lemma collects the comparative statics of the surplus components that enter $D_k$ and therefore $\varXi$.
\begin{lemma}\label{lemma:stw}
    The surplus-to-wage ratio $S_k/w_k$ decreases with the unemployment offer arrival probability $\bar{p}_k^{u}$ and with the separation probability $\gamma_k(w_k)$.
    The marginal surplus $\pd{S_k}{w_k}$ decreases with the on-the-job offer arrival probability $\bar{p}_k^{e}$ and with the separation probability $\gamma_k(w_k)$.
    Moreover, the marginal surplus $\pd{S_k}{w_k}$ increases with $w_k$, and the ratio $S_k/w_k$ increases with $w_k$ provided $\varepsilon_{S_k:w_k}>1$.
\end{lemma}
Lemma \ref{lemma:stw} separates two economically distinct outside-option channels.
Unemployment offers ($\bar p_k^{u}$) primarily affect the \emph{level} of employment rents by raising $U_k$, compressing $S_k$ and therefore $S_k/w_k$.
On-the-job offers ($\bar p_k^{e}$) primarily affect the \emph{marginal} return to a wage increase by making retention less sensitive to the current wage, lowering $\pd{S_k}{w_k}$.
These two components enter the wage wedge through \eqref{eq:varXi}.

\begin{proposition}\label{prop:REF_Xi_behavior}
    For a fixed $\varepsilon_{\E \ell_k:w_k}$, the factor $\varXi$ decreases with the unemployment offer arrival probability $\bar{p}_k^{u}$ and with the separation probability $\gamma_k(w_k)$.
    Moreover, $\varXi$ increases with the wage $w_k$ under the maintained condition $\varepsilon_{S_k:w_k}>1$.
\end{proposition}
Since $\varXi=1/D_k$ and $D_k=1-\eta(1-\beta)\bigl(\frac{S_k}{w_k}\varepsilon_{\E\ell_k:w_k}+\pd{S_k}{w_k}\bigr)$, stronger outside options and higher turnover reduce the rent terms that responsibility internalizes (both in levels and, under our assumptions, in marginal terms), pushing $D_k$ up and therefore shrinking $\varXi$.
Economically, when workers can more easily leave or quickly find jobs after separation, a wage increase generates a smaller welfare gain for incumbents, so responsibility has less bite and the REF behaves closer to a profit-maximizing firm.

\subsection{Perfect competition}
We now examine the limiting behavior of our model as labor supply approaches perfect elasticity, offering a benchmark for understanding the core differences between profit-maximizing and responsible firms. 
This reveals how responsibility alters firm behavior even when market frictions become negligible in terms of wage-setting power.

\begin{proposition}\label{prop:perfect_competition}
    When labor supply approaches perfect elasticity ($\varepsilon_{\E \ell_k : w_k} \to \infty$), the profit-maximizing firm treats the wage rate as fixed at $w_k$, whereas the REF treats it as fixed at $w_k - \eta(1 - \beta)S_k(w_k)$.
\end{proposition}

\begin{figure}[t]
\caption{Use of labor market power by the responsible firm}\label{f:micro_summary}
    \centering
        \begin{tikzpicture}[scale=1, >=latex]
          % Axes
          \draw[->] (0,0) -- (8,0) node[right] {Posted wages};
          \draw[->] (0,0) -- (0,6) node[above] {Outside option / turnover};
          
          % Region boundaries
          \draw[dashed] (4,0) -- (4,6);
          \draw[dashed] (0,3) -- (8,3);
          
          % Annotations
          % Region A
          \node[align=center] at (2,0.5) {\tiny Weak outside option\\[-5pt] \tiny Low Wages};
          \node[align=center] at (2,2) {\small Does \textbf{not} use\\[-1pt] \small market power \\ \scriptsize (eff. cost $\approx$ reservation value)};
          
          % Region B
          \node[align=center] at (6,0.5) {\tiny Weak outside option\\[-5pt] \tiny High Wages};
          \node[align=center] at (6,2) {\small Does \textbf{not} use\\[-1pt] \small market power \\ \scriptsize (eff. cost $\approx$ reservation value)};
          
          % Region C
          \node[align=center] at (2,3.5) {\tiny Strong outside option\\[-5pt] \tiny Low Wages};
          \node[align=center] at (2,5) {\small \textbf{Uses}\\[-1pt] \small market power};
          
          % Region D
          \node[align=center] at (6,3.5) {\tiny Strong outside option\\[-5pt] \tiny High Wages};
          \node[align=center] at (6,5) {\small \textbf{May or may not}\\[-1pt] \small use market power \\[-1pt] \scriptsize (minimal profit gains \\[-5pt] \scriptsize and surplus losses)};

\end{tikzpicture}
\note[Note]{The figure summarizes how responsible firms' use of market power varies across labor market conditions.
The horizontal axis shows the firm's wage relative to the market one, while the vertical axis represents the strength of worker outside options and turnover (high $\bar p_k^{u}$ and $\gamma_k$).
When outside options are weak, REFs refrain from using market power regardless of their wage level. 
When outside options are strong, REFs may exercise market power when offering low wages, as workers face many alternative opportunities. 
When both outside options and wages are high, the use of market power has minimal impact on both profits and worker surplus.}
\end{figure}
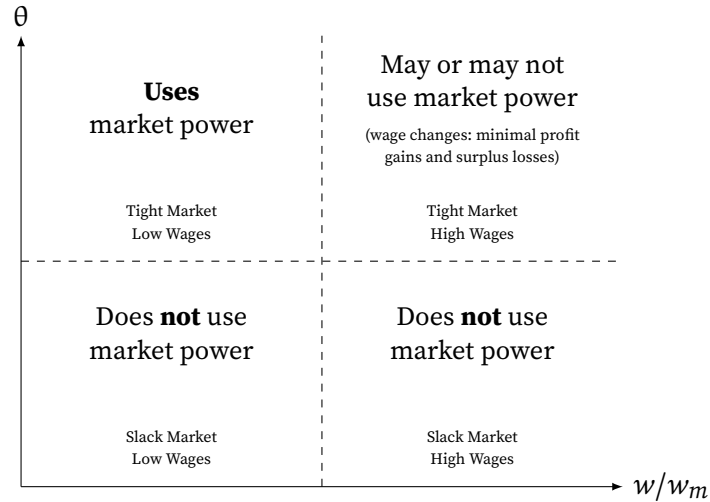

This result represents a significant departure from the benchmark competitive model recalled earlier. 
In conventional theory, the equilibrium wage equals the marginal worker's willingness to accept, determined at either the extensive margin (the reservation wage of the marginal worker) or the intensive margin (the marginal disutility of labor). 
However, in our frictional labor market model, the REF's effective wage benchmark $w_k - \eta(1 - \beta)S_k(w_k)$ lies below the market wage.
This occurs because the firm accounts for the workers' surplus derived from employment stability in environments with search frictions, effectively reducing its perceived labor costs relative to the nominal wage rate.

Our findings are summarized in Figure \ref{f:micro_summary} and substantially alter the standard result that a responsible firm must price labor competitively everywhere: here, the implications of responsibility for wage-posting are contingent on market frictions and involve the reservation value below the market wage.

%%%%%%%%%%%%%%%%%%%%%%%%%%%%%%%%%%%%%%%%%%%%%%%%%%%%%%%%%%%%%%%%%%
% SECTION: BM MODEL WITH RESPONSIBLE FIRMS
%%%%%%%%%%%%%%%%%%%%%%%%%%%%%%%%%%%%%%%%%%%%%%%%%%%%%%%%%%%%%%%%%%
\section{A wage-posting model with responsible firms}\label{sec:BM}
Section~\ref{sec:micro_model} provided a reduced-form, firm-level formulation in which responsibility enters as a direct valuation of worker welfare, allowing us to highlight how outside options and turnover shape responsible wage and vacancy choices. 
We now shift to a market-level perspective in which wages are posted and workers search both while unemployed and on the job, so the equilibrium object of interest is the endogenous wage distribution.
Throughout, responsibility continues to mean that the firm places extra weight on workers' utility; what changes is the environment through which this concern maps into wages, employment, and sorting.

We build a wage-posting model with on-the-job search in which profit-maximizing firms coexist with responsible firms that value worker welfare.
The main object of interest is the endogenous wage distribution---and how responsibility and productivity shape the sorting of firm types across it.

\subsection{Baseline: wage posting with efficiency wages}
We start from the search frictions of \cite{burdett1998wage} (hereafter BM) and add a simple efficiency-wage channel: posted wages affect effort and hence output.
This baseline is the benchmark before introducing responsibility.

\paragraph{Time and payoffs}
We follow the standard BM formulation and work in continuous time: job offers arrive at Poisson rate $\lambda$ and matches dissolve at rate $\delta$, so the equilibrium conditions are steady-state flow equalities.\footnote{%
This contrasts with the discrete-time notation used elsewhere in the paper; one can interpret $\lambda$ and $\delta$ as per-period transition probabilities (or as rates over a short period length).}

\paragraph{Environment}
The economy features a mass 1 of equally productive workers, each with a non-work option $b$, and a continuum of firms, with mass 1 and $T$ types.
A firm of type $i$ has production technology $y_i(e)$, where $e$ is the level of effort put by workers on the job.
Exerting effort incurs disutility $c(e,w)$, with $c_{ew}<0$ capturing that higher wages reduce the marginal disutility of effort (motivation).\footnote{%
This reduced-form specification can capture reciprocity, morale, identity, or other psychological channels.}
Workers do not internalize output; the efficiency-wage channel operates because wages shift the effort cost schedule.
A firm offers a uniform wage $w$ to all its workers (which may differ among firms of the same type) to maximize steady-state profits. 
Workers—whether employed or not—receive job offers randomly at an exogenous rate $\lambda$, drawn from the endogenous wage distribution $F(w)$. 
Unemployed workers accept offers above the reservation wage $w^r$, while employed workers accept offers exceeding their current wage.\footnote{%
The reservation wage $w^r$ is pinned down by the outside option $b$ via $w^r-c(e^*(w^r),w^r)=b$, where $e^*(w)$ is the worker’s optimal effort at wage $w$.
We assume no mass point at $w^r$, so $F(w^r)=0$.
}
Matches dissolve at rate $\delta$.

We will work with $T=2$ firm types.
In the baseline analysis we take $y_2(e) > y_1(e)$ for all $e$ for ease of exposition.
When introducing responsible firms below we explicitly distinguish the three cases $y_2 = y_1$, $y_2 > y_1$, and $y_2 < y_1$.
The prevailing distribution of posted wages is a mix of the distributions for the two types:
\[
    F(w) = \alpha F_1(w) + (1-\alpha)F_2(w)
\]
where $\alpha\in(0,1)$ is the population share of type--1 firms. 
Our goal is to recover the equilibrium distributions and profits by type, $\bc{F_1, F_2, \pi_1, \pi_2}$, and the overall wage distribution $F$.

\begin{assumption}\label{ass:eprime}
    The cost function is convex in $e$ ($\frac{\partial^2 c}{\partial e^2} > 0$) and  the marginal cost of effort decreases with the wage ($\frac{\partial^2 c}{\partial e \partial w} < 0$). We further assume that $\frac{\partial c}{\partial w} < 1$.
\end{assumption}

Given a posted wage $w$, workers choose effort to maximize $w-c(e,w)$.
Effort does not raise the worker's pay directly; instead, a higher wage shifts the disutility of effort (``gift exchange''/motivation), captured by $c_{ew}(e,w)<0$.
Let $e^*(w)$ denote the worker's optimal effort. Under the maintained interiority assumption, it is characterized by $c_e(e^*(w),w)=0$.
Define the worker's net flow utility from a job paying wage $w$ as
\[
    \tilde u(w):=w-c(e^*(w),w),
\]
and recall that the reservation wage $w^r$ satisfies $\tilde u(w^r)=b$.
We define flow surplus relative to the outside option as
\[
    S(w):=\tilde u(w)-b,
\]
so $S(w^r)=0$ and $S(w)\ge 0$ for any wage $w\ge w^r$ that workers accept.
By Assumption \ref{ass:eprime},
\[
    e^{*\prime}(w)=- \frac{c_{ew}(e^*(w),w)}{c_{ee}(e^*(w),w)} > 0,
\]
and the surplus is increasing in the wage because
\[
    S'(w)=\frac{d}{dw}\bs{\tilde u(w)-b}=1-c_w(e^*(w),w)>0.
\]

\begin{assumption}\label{ass:ydiff}
    Type-2  technology is always more responsive to worker effort than type-1  technology:
    \[ y_2'(e) \ge y_1'(e) \quad \text{for all } e. \]
\end{assumption}

The steady-state equilibrium is determined by the four conditions discussed next.

\paragraph{Unemployment condition}
The proportion of unemployed workers is $u$, with the flow of workers into unemployment given by $\delta (1-u)$.
The flow of workers out of unemployment is $\lambda \bs{1-F(w^r)} u$.
Equating the two flows yields the unemployment rate
\[
    u = \frac{\delta}{\delta + \lambda}
\]
with $F(w^r)=0$, since no firm will offer a wage below the reservation wage in equilibrium.

\paragraph{Workers' flows condition}
Let $G(w)$ be the proportion of employed workers earning a wage no greater than $w$. 
The size of the pool of workers in these jobs is $(1-u) G(w)$.
The net entry to this set comes from the pool of unemployed workers.
The inflow is $u \lambda F(w)$ since a share $\lambda F(w)$ of the unemployed receives offers between $w^r$ and $w$.
The outflow consists of two parts: \emph{a)} the job destruction flow, $\delta (1-u) G(w)$, and \emph{b)} the exit to higher-wage jobs, $\lambda (1-u) G(w) \bp{1-F(w)}$.
Equating inflow and outflow, we get the condition
\[
    u \lambda F(w) = \delta (1-u) G(w) + \lambda (1-u) G(w) \bp{1-F(w)}
\]
From this, we can derive the steady-state distribution
\begin{equation*}
    G(w) = \frac{u \lambda F(w)}{(1-u) \bs{\delta + \lambda \bp{1-F(w)}}}
    = \frac{ F(w)}{1 + \frac{\lambda}{\delta} \bp{1-F(w)}} < F(w)
\end{equation*}
This classical result shows that the distribution of paid wages  first-order stochastically dominates the distribution of offered wages.

\paragraph{Profits condition} 
Every match between a worker and a firm of type $i$ posting wage $w$ generates a productivity $y_i(e^*(w))$ and costs $w$ to the firm.
Denoting $\ell(w)$ the measure of workers per firm posting a wage $w$, the profit of a type $i$ firm posting wage $w$ is $\pi_i = \ell(w) \bs{y_i(e^*(w)) - w}$.
In a mixed-strategy equilibrium, all active firms of a given type earn the same profit, so $\pi_i$ is constant on the support of $F_i$.
Under $y_2(e) > y_1(e)$ and Assumption \ref{ass:ydiff}, equilibrium is stratified: type--2 firms post strictly higher wages than type--1 firms, so the supports of $F_1$ and $F_2$ are ordered and disjoint.\footnote{%
A convenient way to see the ordering is to compare deviations. Fix $w_1$ in the support of $F_1$ and $w_2$ in the support of $F_2$.
Value constancy gives $\ell(w_2)\bs{y_2(e^*(w_2))-w_2}\ge \ell(w_1)\bs{y_2(e^*(w_1))-w_1}>\ell(w_1)\bs{y_1(e^*(w_1))-w_1}$ and $\ell(w_1)\bs{y_1(e^*(w_1))-w_1}\ge \ell(w_2)\bs{y_1(e^*(w_2))-w_2}$.
Combining yields $\ell(w_2)\big(y_2(e^*(w_2))-y_1(e^*(w_2))\big)>\ell(w_1)\big(y_2(e^*(w_1))-y_1(e^*(w_1))\big)$ and hence $w_2>w_1$ because $\ell$ and $e^*$ are increasing and $y_2-y_1$ is non-decreasing.}

Firm size at wage $w$ equals the share of employed workers at $w$ divided by the density of posted offers:\footnote{%
When supports are disjoint, this common $\ell(w)$ specializes piecewise to each type’s support.
}
\[
    \ell(w) = \lim_{\varepsilon \to 0} \frac{G(w) - G(w-\varepsilon)}{F(w) - F(w-\varepsilon)} (1-u) = \frac{g(w)}{f(w)} (1-u)
\]
where $g(w) = G'(w) = \frac{\delta (\delta+\lambda) f(w)}{\bs{\delta + \lambda \bp{1-F(w)}}^2}$.
Substituting this expression back into $\ell(w)$ yields the two profit functions $\pi_i = \ell(w)\bs{y_i(e^*(w)) - w}$:
\[
    \pi_1 = \frac{\delta\lambda \bs{y_1(e^*(w)) - w}}{\bs{\delta + \lambda \bp{1- \a F_1(w)}}^2} \qquad
    \pi_2 = \frac{\delta\lambda \bs{y_2(e^*(w)) - w}}{\bs{\delta + \lambda (1-\a) \bp{1-  F_2(w)}}^2}
\]

\paragraph{Equilibrium wage offers condition}
Let $\operatorname{Supp}(F_i)=[\underline w_i,\bar w_i]$ denote the support of type $i$'s (conditional) wage distribution.
In the stratified equilibrium described above, $\underline w_1=w^r$ and $\bar w_1=\underline w_2=:w_b$, so $F(w^r)=0$ and $F(w_b)=\alpha$.
Since profit is uniform in a given type, it is equal to the lowest-wage firm's profit:
\[
    \pi_i=
    \frac{\delta\lambda \bs{y_i(e^*(w))-w}}{\bs{\delta+\lambda(1-F(w))}^2}
    =\frac{\delta\lambda \bs{y_i(e^*(\underline w_i))-\underline w_i}}{\bs{\delta+\lambda(1-F(\underline w_i))}^2}
\]
Using $F(\ul{w}_1)=0$ and $F(\ul{w}_2)=\a$ (i.e.\ $F(w^r)=0$ and $F(w_b)=\a$), we can recover $F_i(w)$ for each $i=1,2$ from the mixed wage distribution:\footnote{%
The objects $F_1$ and $F_2$ here are \emph{conditional} cumulative distribution functions: each integrates to 1 over the support of its own firm‑type.
The aggregate wage distribution is therefore $F(w)= \alpha F_1(w) + (1-\alpha) F_2(w)$, where $\alpha$ is the population share of type‑1 firms.
Several presentations in the literature absorb the mixing weights into the definitions of $F_1$ and $F_2$. 
In that convention the factors $1/\alpha$ and $1/(1-\alpha)$ in \eqref{eq:BM_wagedist} disappear, but the mixed distribution $F(w)$ is, of course, identical.
}
\begin{equation}\label{eq:BM_wagedist}
    \begin{split}
    F_1(w) &= \frac{1}{\a} F(w) = \frac{\delta +\lambda}{\lambda\a} \bp{1 - \sqrt{\frac{y_1(e^*(w)) - w}{y_1(e^*(\ul{w}_1)) - \ul{w}_1}}} \\
    F_2(w) &= \frac{F(w)-\a}{1-\a} = \frac{\delta +\lambda (1-\a)}{\lambda (1-\a)} \bp{1 - \sqrt{\frac{y_2(e^*(w)) - w}{y_2(e^*(\ul{w}_2)) - \ul{w}_2}}}
\end{split}
\end{equation}
which induces the general wage distribution $F(w)=\alpha F_1(w) + (1-\alpha)F_2(w)$ (where $F_i(w)=0$ or $1$ when $w$ is outside its support). 
Since $\ul{w}_1=w^r$, $\bar{w}_1 = \ul{w}_2$, and $F_1(\bar{w}_1)=1$:
\begin{equation*}
    y_1(e^*(\bar{w}_1)) - \bar{w}_1 = \bs{y_1(e^*(w^r)) - w^r} \bp{\frac{\delta + \lambda (1 - \a)}{\delta + \lambda}}^2
\end{equation*}
and the profit levels are therefore
\[
    \pi_1 = \frac{\delta \lambda}{(\delta + \lambda)^2} \bs{y_1(e^*(w^r)) - w^r}, \quad 
    \pi_2 = \frac{\delta \lambda}{\bs{\delta + \lambda(1-\a)}^2} \bs{y_2(e^*(\ul{w}_2)) - \ul{w}_2}
\]

\subsection{Responsible firms}
We now allow type-2 firms to partially internalize incumbent-worker \emph{flow} surplus. 
A type-$i$ firm chooses a posted wage $w$ to maximize
\[
    \max_{w} \ell (w) \Bigl[ y_i (e^*(w)) - w + \ind{i=2} \, \eta \bs{\tilde u(w) - b} \Bigr]
\]
where $\eta\in[0,1)$ is the type-2 (responsible) firm's responsibility degree.
Here the internalized term is incumbent-worker \emph{flow} welfare: the firm values the contemporaneous net utility of its employed workers at wage $w$, relative to the outside option $b$ (so $S(w^r)=0$).\footnote{%
This differs from the present-value search surplus $W(w)-U$ used in earlier dynamic formulations.
Internalizing that PV object in a BM wage-posting environment would require tracking continuation values under on-the-job search and the endogenous wage distribution; the flow formulation is a reduced-form proxy that preserves the key monotonicities while keeping the wage-posting problem tractable.}
To avoid confusion with the unemployment rate $u$, recall $\tilde u(w)=w-c(e^*(w),w)$ and $S(w)=\tilde u(w)-b$, with $w^r$ defined by $\tilde u(w^r)=b$.
Throughout,
\[
    \ell(w) := \frac{\lambda\delta}{\big[\delta + \lambda(1-F(w))\big]^{2}}
\]
is the standard Burdett–Mortensen size mapping, obtained from steady-state flow balance.

Let $e^*(w)$ be the worker’s optimal effort at wage $w$.
Define the per-worker profit margins
\[
    M_1(w) = Y_1(w), 
    \qquad
    M_2(w) = Y_2(w) + \eta S(w)
\]
with $Y_i(w) = y_i (e^*(w)) - w$ and $S(w) = \tilde u(w) - b$. 
Then
\[
    Y_i'(w) = y_i' (e^*(w)) e^{* \prime}(w) - 1, \qquad
    M_2'(w)=Y_2'(w) + \eta S'(w)
\]
where
\[
    S'(w) = \frac{d}{dw} \bs{\tilde u(w)-b}
    = 1 - c_w (e^*(w), w ) 
\]
is the marginal surplus (using the worker’s FOC $c_e(e^*(w),w)=0$ and the envelope theorem).

\begin{assumption}\label{ass:BM_decreasing_margin}
    For each type $i$, the per-worker cash margin $Y_i(w)=y_i(e^*(w))-w$ is strictly decreasing on any wage interval used in equilibrium:
    \[
    Y_i'(w)=y_i'(e^*(w))\,e^{*\prime}(w)-1<0.
    \]
\end{assumption}
This is a mild regularity condition: it rules out regions where the induced effort increase from a higher wage raises cash output by more than one-for-one, so that higher wages would mechanically increase per-worker cash profits.
Under the linear specification $y_i(e)=y_{i,0}+\xi_i e$ and $e^*(w)=(\eta_c/k)w$, it reduces to $\xi_i(\eta_c/k)<1$.

We now study three cases: \emph{i)} REFs and PMFs have the same productivity; \emph{ii)} REFs are more productive; \emph{iii)} PMFs are more productive.

\begin{proposition}\label{prop:BM_equal_prod}
    If $y_2(e) = y_1(e)$ for all $e$, then no responsible firm posts a wage strictly below a wage posted by a profit--maximizing firm.
    In particular, REFs (type 2) occupy the upper wage segment and PMFs (type 1) occupy the lower one.
\end{proposition}

With identical productivities, REFs have strictly higher per-worker margins at any wage $w>w^r$ because they internalize $S(w)=\tilde u(w)-b\ge 0$ (with strict inequality for wages strictly above the reservation level). 
Since the Burdett–Mortensen size effect $\ell(w)$ rises with the wage, REFs always prefer higher wages relative to PMFs. 
Thus PMFs cannot appear strictly above REFs in the wage distribution; when supports are separated, this yields the segment ordering stated in Proposition \ref{prop:BM_equal_prod}.

\begin{proposition}\label{prop:BM_REFs_mp}
    If $y_2(e) > y_1(e)$ and $y_2'(e) \ge y_1'(e)$ for all $e$, then no responsible firm posts a wage strictly below a wage posted by a profit--maximizing firm.
    In particular, REFs (type 2) occupy the upper wage segment and PMFs (type 1) occupy the lower one.
\end{proposition}

Higher productivity pushes REFs toward higher wages even without responsibility. 
Adding $\eta > 0$ reinforces this upward pull through surplus internalization. 
The size effect $\ell(w)$ amplifies both forces. 
Thus the only consistent ordering is PMFs lower, REFs higher.

Next, we give a sufficient condition ensuring that the ``segment'' language above is literal, i.e.\ the wage supports of the two types are disjoint (except possibly at a single boundary point).

\begin{proposition}\label{prop:BM_disjoint}
    Assume $\eta>0$ and, on any active interval, $Y_1''(w)=Y_2''(w)\equiv 0$ while $S''(w)\not\equiv 0$.
    Supports $\operatorname{Supp}(F_1)$ and $\operatorname{Supp}(F_2)$ are disjoint in any non-degenerate equilibrium, and can meet at most at a single boundary wage $w_b$, where $F(w_b)=\alpha$ or $F(w_b)=1-\alpha$ depending on which type is lower.
\end{proposition}

We use the assumption of linear technology just to establish the result on supports separation, but make no further use of it in the rest of this discussion, where we simply assume separation holds.

\begin{proposition}\label{prop:BM_REFs_lp}
    Suppose $y_2(e)<y_1(e)$ for all $e$. 
    i) In any equilibrium in which REFs occupy the higher wage segment, one must have
    \[
    \eta \geq \eta^*
    =
    \frac{r_\alpha Y_2(w^r) - Y_2(w_b)}
    {S(w_b)}
    \]
    where $w_b$ solves
    \[
    \frac{Y_1(w_b)}{Y_1(w^r)} = r_\alpha
    =
    \left(\frac{\delta+\lambda(1-\alpha)}{\delta+\lambda}\right)^2.
    \]
    ii) In any equilibrium in which REFs occupy the lower wage segment, one must have
    \[
    \eta \leq \eta^{**}
    =
    \frac{r^*_\alpha Y_2(w^*_b) - Y_2(\bar{w})}
    {S(\bar{w}) - r^*_\alpha S(w^*_b)}
    \]
    where $F(w^*_b)=1-\alpha$, $F(\bar{w})=1$, and $w^*_b, \bar{w}$ solve
    \[
    \frac{Y_1(\bar{w})}{Y_1(w^*_b)} = r^*_\alpha
    =
    \left(\frac{\delta}{\delta+\lambda\alpha}\right)^2.
    \]
\end{proposition}

Here, $r_\alpha\in(0,1)$ is the size wedge between the bottom wage $w^r$ and the boundary $w_b$ when PMFs form the lower segment (case~(i)).
In the REF-lower configuration (case~(ii)), the analogous wedge is $r^*_\alpha\in(0,1)$, defined between the boundary $w^*_b$ and the upper endpoint $\bar w$.
In case~(i), the numerator $r_\alpha Y_2(w^r)-Y_2(w_b)$ is the productivity-margin disadvantage of REFs at the boundary (relative to the bottom, scaled by the size wedge), and the denominator is the flow surplus they internalize at the boundary, $S(w_b)=\tilde u(w_b)-b$ (with $S(w^r)=0$ by construction).
Overall, $\eta^*$ is the minimal responsibility degree making the internalized surplus large enough to offset the productivity disadvantage over the lower segment.
In case~(ii), the analogous interpretation applies after replacing $r_\alpha$ by $r^*_\alpha$ and evaluating the objects at the relevant endpoints $(w^*_b,\bar w)$.

Note that, since $Y_i'(w) = y_i'(e^*(w)) e^{*\prime}(w) - 1$ and $M_2'(w) = Y_2'(w) + \eta \bs{1 - c_w(e^*(w),w)}$, a higher marginal productivity of effort $y_2'(e)$, or a stronger effort response $e^{*\prime}(w)$, raises $Y_2(w_b)$ and shifts the numerator of $\eta^*$ down, reducing the threshold $\eta^*$ and making REF-upper assignments easier to sustain.

\begin{figure}[p]
    \caption{Responsibility, productivity, and support ordering}
    \label{fig:resp_eta_switch}
    \includegraphics[scale=0.56]
    {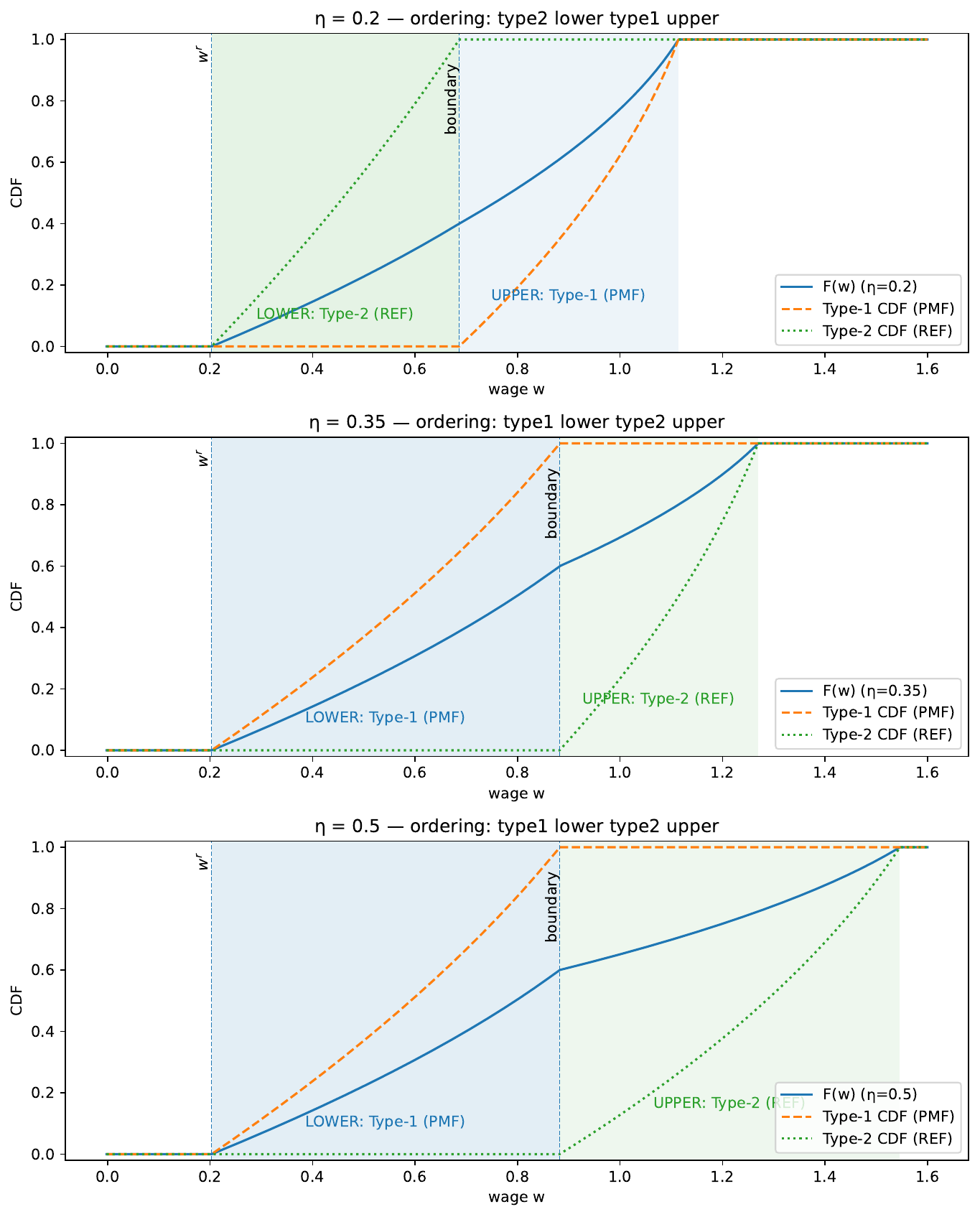}
    \note[Note]{
    Each panel shows the equilibrium wage distribution $F(w)$ (solid) and the conditional CDFs of profit–maximizing firms (Type~1, dashed) and responsible firms (Type~2, dotted) at responsibility levels $\eta \in \{0.2,0.35,0.5\}$.
    Shaded regions indicate which type occupies lower vs upper wage segments; the vertical line marks the boundary $w_b$.
    Type~2 firms are less productive:
    $y_1(e)= y_{1,0} + \xi_1 e$ and $y_2(e)= y_{2,0} + \xi_2 e$ with
    $y_{1,0}=1.0$, $y_{2,0}=0.8$, and $\xi_1 = \xi_2=0.25$.
    Effort is $e^*(w) = (\eta_c/k)w$ with $k=1$, $\eta_c=0.6$, and effort cost
    $c(e,w) = \tfrac{k}{2}e^2 - \eta_c w e + \tfrac{\chi}{2}w^2$ with $\chi=0.5$.
    Moreover, $\delta=0.10$, $\lambda=0.30$, and $\alpha=0.60$.
    For $\eta=0.2$ (top panel), responsible firms occupy the lower segment and
    profit–maximizers the upper segment.
    For $\eta=0.35$ (middle panel), the ordering switches: profit–maximizers are
    lower and responsible firms are upper.
    For $\eta=0.5$ (bottom panel), the ordering remains PMF–lower/REF–upper,
    and the responsible firms’ support extends further to the right, showing how higher responsibility pushes responsible firms into the high–wage tail despite their productivity disadvantage.
    }
\end{figure}

Figure \ref{fig:resp_eta_switch} provides a numerical illustration of the case in which type-2 firms are less productive.
When responsibility is low ($\eta=0.2$), responsible firms lie on the lower mixed segment and profit-maximizers on the upper one.
Once $\eta$ exceeds the analytical threshold $\eta^*$ (given the parameterization, for $\eta=0.35, 0.5$ in the Figure), the ordering switches: profit-maximizers become the lower type and responsible firms form the upper tail of the wage distribution.
The figure shows the implied boundary $w_b$ and the evolution of both supports as $\eta$ increases.

\paragraph{Full responsibility}
With full responsibility, the firm fully internalizes incumbent-worker flow welfare, so the objective becomes total match surplus (output net of effort disutility, minus the outside option) multiplied by firm size.
We impose a solvency constraint requiring nonnegative cash profits, so that firms do not choose wages that exceed contemporaneous output and would imply exit.
With $\eta=1$, type–2 firms solve
\[
    \max_w \ell(w) \bs{y_2(e^*(w)) - c(e^*(w),w) - b}
    \quad \text{s.t.} \quad
    \ell(w) \bs{y_2(e^*(w)) - w} \ge 0
\]
Since $\ell(w)>0$, the constraint is $y_2(e^*(w))\ge w$.
Define
\[
    \hat w \in \arg\max_w \ell(w) \bs{y_2(e^*(w)) - c(e^*(w),w) - b},
    \qquad
    w^* := \sup\{w: y_2(e^*(w)) \ge w\}
\]
Assume the objective is single–peaked and that $y_2(e^*(w))$ crosses the $45^\circ$ line at most once.
The constrained optimum is therefore
\[
    w^\star_{FR} = \min\{\hat w, w^* \}
\]
All fully responsible firms select $w^\star_{FR}$, so type–2 wages bunch at a single point.

If, in addition, effort is constant ($e^*(w)\equiv \bar e$) and $c\equiv 0$, then
\[
    \ell(w) \bs{y_2(\bar e)-b} \text{ is increasing in $w$},\qquad
    w^*=y_2(\bar e)
\]
hence $w^\star_{FR}=y_2(\bar e)$.

\subsection{Discussion}
This section extends the Burdett--Mortensen wage–posting framework by allowing some firms to internalize worker surplus and by making productivity wage-dependent through effort.
Responsible firms (type~2) give weight $\eta$ to worker surplus, while profit–maximizing firms (type~1) do not.
The combination of on–the–job search, efficiency wages, and responsibility generates a rich but tractable structure for the equilibrium wage distribution.

In a separated-support equilibrium, wages can be partitioned into a lower segment and an upper segment, each occupied by a single firm type.
Propositions \ref{prop:BM_equal_prod}--\ref{prop:BM_REFs_mp} establish the ordering: when responsible firms are at least as productive as profit--maximizers, they cannot lie below them in the wage distribution.
Proposition \ref{prop:BM_disjoint} provides a sufficient condition ensuring that the supports are in fact disjoint (except possibly at a single boundary point).
The question is then: \emph{which} type sits in which segment, and how does this depend on productivity and the responsibility degree?

When responsible firms are at least as productive as profit–maximizers, the ordering is unambiguous.
If technologies are identical, responsible firms always earn a strictly higher margin at any given wage because they internalize worker surplus.
Since the size mapping $\ell(w)$ and the worker surplus $S(w)$ are both increasing in the wage, responsible firms are strictly more attracted to high wages than profit–maximizers.
They therefore occupy the upper part of the wage distribution, with profit–maximizers forming the lower part.
When type-2 firms are also more productive, this upward pull is reinforced, provided that the marginal productivity of effort is not lower for these firms: then, both the productive component and the surplus component of the margin favor higher wages for responsible firms.
In both cases, responsibility is associated with higher wages and stronger efficiency–wage effects in steady state.

The most interesting case is when responsible firms are \emph{less} productive.
Here responsibility and productivity work in opposite directions.
On the one hand, profit–maximizers have a higher productive margin for any given effort level.
On the other hand, responsible firms value worker surplus and therefore benefit more from raising wages, especially where the marginal surplus $S(w)$ is large.
The comparison is governed by the value of $\eta$.
When $\eta $ is low enough, the productivity disadvantage dominates: responsible firms take the lower wage segment and undercut profit–maximizers.
Workers who end up at responsible firms receive low wages, because these firms cannot afford to “spend” much on surplus internalization.
When $\eta $ is high enough, responsibility dominates: profit–maximizers form the lower segment, while responsible firms become the high–wage employers despite their lower productivity.
In this region, responsible firms willingly sacrifice profits to internalize enough worker surplus---this triggers stronger effort responses, partially offsetting their technological disadvantage.

In the limit case of full responsibility, type–2 firms maximize match surplus (output net of effort disutility) subject to solvency.
All responsible firms optimally choose the same wage, so their offers bunch at a single point.
Partial responsibility thus generates dispersion and sorting across wage segments, while full responsibility leads to wage compression within the responsible–firm segment.

%%%%%%%%%%%%%%%%%%%%%%%%%%%%%%%%%%%%%%%%%%%%%%%%%%%%%%%%%%%
% % DMP model with responsible firms
%%%%%%%%%%%%%%%%%%%%%%%%%%%%%%%%%%%%%%%%%%%%%%%%%%%%%%%%%%%
\section{A DMP model with responsible firms}\label{sec:DMP}
The wage-posting framework of Section~\ref{sec:BM} is well suited to studying wage dispersion and sorting, but it abstracts from vacancy creation and from bargaining inside matches. 
We therefore close with a canonical search-and-matching model with endogenous tightness, where wages are pinned down by a bargaining protocol.\footnote{%
This DMP environment abstracts from wage dispersion: in the stationary symmetric benchmark with Nash bargaining, the wage is uniquely determined by the bargaining rule given equilibrium objects. 
Wage dispersion would re-emerge under wage posting, heterogeneity or directed search, but those ingredients are not needed for the general-equilibrium tightness and governance mechanisms highlighted in this section.}
In that environment, there is no direct wage policy inside the match beyond the bargaining rule itself, so we model responsibility as a commitment to a more worker-friendly protocol (a higher worker bargaining weight), which preserves the interpretation of responsibility as giving workers a larger share of match surplus.
This section thus develops a Diamond--Mortensen--Pissarides (DMP) model in which firms can be ``responsible'' in their wage-setting.

We proceed in three steps.
First, we summarize the standard equilibrium under free and limited entry.
Second, we study the normative question of ``optimal governance'': what worker bargaining weight maximizes stakeholder value once equilibrium market tightness adjusts.
Third, we turn to a positive exercise in which only a subset of firms commits to higher worker bargaining power, and we characterize wage spillovers on regular firms.

\subsection{The standard DMP setting}

\paragraph{Environment}
Time is discrete and infinite.
Workers receive an unemployment value $b$ when not employed.
Firms each employ at most one worker and produce $y$ when matched.
An unmatched firm posts a vacancy at per-period cost $\kappa$; we abstract from an additional ``idle'' margin, so the stock of firms is split between filled jobs and vacancies.

Once a vacancy is filled, the match produces $y$ units of output and incurs a wage cost $w$. 
In each period, the match dissolves with exogenous probability $\delta$. 
Wages result from an endogenous bargaining process
\[
    w^* \equiv \argmax_w S_\omega^\phi S_f^{1-\phi}
\]
where $\phi$ is the worker's bargaining power, and $S_\omega$ and $S_f$ are the worker's and firm's surpluses, respectively.
The bargaining stage takes $(U,V,\theta)$ as given: bargaining determines the wage conditional on current outside values and tightness, while those objects are pinned down in equilibrium.

The matching process follows the function $M(u,v)$, where $u$ and $v$ are the aggregate measures of unemployed workers and vacancies, respectively.
The matching function is concave and increasing in both arguments. 
Labor market tightness is $\theta = v/u$. 
Given $M(u,v)$ is homogeneous of degree one, the probability of filling a vacancy is $q(\theta) = M(1/\theta, 1)$, and the probability an unemployed worker finds a job is $p(\theta) = \theta q(\theta)$. 
Since $M(u,v)$ is increasing, $q(\theta)$ is decreasing, and $p(\theta)$ is increasing in $\theta$.

Both workers and firms discount the future at a common discount rate $\beta < 1$. 
A representative worker can be either employed or unemployed at any given time. 
In stationary equilibrium, the present values of an employed or unemployed worker are respectively given by
\[
    W = w + \beta \bs{ \delta U + (1-\delta)W} \quad \text{and} \quad U = b + \beta \bs{\theta q(\theta)W + (1- \theta q(\theta)) U }.
\]
Similarly, a firm with a filled or unfilled vacancy has value
\[
    J = y -w + \beta \bs{\delta V + (1-\delta) J}
    \quad \text{and} \quad
    V = -\kappa + \beta\bs{q(\theta)J + (1-q(\theta))V}.
\]

The laws of motion establish steady-state relationships between unemployment, vacancies, and market parameters: $u = \frac{\delta}{\delta + \theta q(\theta)}$ and $v = \frac{\delta \theta}{\delta + \theta q(\theta)}$.
The total number of firms in the economy is $N = v + (1-u)$.
In a steady state, this becomes
\begin{equation}\label{eq:DMP_Nfirms}
    N = \theta \frac{\delta + q(\theta)}{\delta + \theta q(\theta)}.
\end{equation}
In particular, on slack markets ($\theta\le 1$) the right-hand side is strictly increasing in $\theta$ whenever the job-finding probability is increasing in $\theta$, as under standard matching technologies (e.g., Cobb--Douglas), so \eqref{eq:DMP_Nfirms} can be inverted to map $N$ into an equilibrium tightness $\theta$.

\paragraph{Surplus and bargaining}
Worker and firm surpluses are
\[
    S_\omega = W-U = \frac{w-b}{1 - \beta \bp{1-\delta-\theta q(\theta)}}
    \quad \text{and} \quad
    S_f = J-V = \frac{y-w+\kappa}{1 - \beta \bp{1-\delta- q(\theta)}}
\]
and the Nash-bargained wage is characterized by
\begin{equation}\label{eq:DMP_wage_bargaining}
    w = \phi\bp{y-V(1-\beta)}+(1-\phi)U(1-\beta)
\end{equation}
where $U$ and $V$ (and hence $\theta$) are taken as given at the bargaining stage.
We provide the algebraic derivations of these expressions in Appendix~\ref{a:add_computations_DMP}.

While much of the literature assumes free entry of firms, we also examine the case where free entry does not hold. 
This is relevant in markets where firms with varying levels of responsibility coexist.

\paragraph{Free entry equilibrium}
Under free entry, the number of firms $N$ is endogenous. 
Free entry drives the value of an open vacancy to zero, so $V=0$.
Using the vacancy value equation then gives $\kappa = \beta q(\theta)J$, which implies the job-creation curve:
\begin{equation}\label{eq:DMP_beveridge}
    w = y - \kappa \frac{1-\beta(1-\delta)}{\beta q(\theta)}    
\end{equation}
which shows a negative relation between $w$ and $\theta$, as $q(\theta)$ decreases in $\theta$.

On the wage-setting side, Nash bargaining implies an increasing wage curve,
\begin{equation}\label{eq:DMP_eqw_FE}
    w = \varphi(\theta) y + (1-\varphi(\theta)) b
\end{equation}
where
\[
    \varphi(\theta) = \frac{\bs{1-\beta \bp{1-\delta-\theta q(\theta)}}\phi}{\bs{1-\beta \bp{1-\delta-\theta q(\theta)}}\phi + \bs{1-\beta(1-\delta)}(1-\phi)}.
\]
This function is increasing in both $\phi$ and $\theta$.
Equilibrium tightness is pinned down by the intersection of the decreasing job-creation curve \eqref{eq:DMP_beveridge} and the increasing wage-setting curve \eqref{eq:DMP_eqw_FE}, i.e.
\begin{equation}\label{eq:DMP_tightness_condition}
    \varphi(\theta) y + (1-\varphi(\theta)) b = 
    y - \kappa \frac{1-\beta(1-\delta)}{\beta q(\theta)}
\end{equation}
The left-hand side increases in $\theta$, while the right-hand side decreases, implying a unique solution for $\theta$.
In particular, holding the matching technology fixed, higher bargaining power shifts the wage-setting curve up and lowers equilibrium tightness.

\paragraph{Equilibrium with no free entry}
Without free entry, the number of firms $N$ is fixed and, given that all unmatched firms post vacancies, market tightness is pinned down by the steady-state relationship \eqref{eq:DMP_Nfirms}.
Since entry is restricted, the vacancy value $V$ need not be zero.
The bargaining wage can be written as
\begin{equation}\label{eq:DMP_eqw_NFE}
    w = \hat{\varphi}(\theta)(y + \kappa) + (1-\hat{\varphi}(\theta))b
\end{equation}
where $\hat{\varphi}(\theta)$ is defined as:
\[
    \hat{\varphi}(\theta) = \frac{\phi\bs{1-\beta\bp{1-\delta-\theta q(\theta)}}}{\phi \bs{1-\beta \bp{1-\delta-\theta q(\theta)}} + (1-\phi) \bs{1-\beta \bp{1-\delta-q(\theta)}}}
\]
This function $\hat{\varphi}(\theta)$ shares the same properties as $\varphi(\theta)$ described in the case of free entry.

\paragraph{Elastic labor supply}
To connect governance to participation, we also consider an extensive-margin labor supply $L(w)$, with $L(w)=0$ for $w\le b$.

The steady-state variables become $u = \frac{\delta L(w)}{\delta + \theta q(\theta)}$ and $v = \frac{\delta \theta L(w)}{\delta + \theta q(\theta)}$, while the number of firms is:
\begin{equation}\label{eq:DMP_els_Nfirms}
    N = v+\left(L(w)-u\right)
    = L(w) \theta \frac{\delta + q(\theta)}{\delta + \theta q(\theta)}.
\end{equation}%

With free entry, $(w,\theta)$ are pinned down by the job-creation and wage-setting conditions \eqref{eq:DMP_beveridge}--\eqref{eq:DMP_tightness_condition} and therefore do not depend on $L(\cdot)$.
Given the equilibrium wage, labor supply scales the equilibrium measures through \eqref{eq:DMP_els_Nfirms} (and the associated steady-state relationships for employment and unemployment).

Without free entry, market tightness depends on the wage and decreases as $w$ rises since a higher wage attracts more workers, reducing tightness. 
Thus, $w$ and $\theta$ are jointly determined by \eqref{eq:DMP_eqw_NFE} and \eqref{eq:DMP_els_Nfirms}, where the first equation shows a positive relation between $w$ and $\theta$, while the second shows a negative relation.

\subsection{Seeking optimal corporate governance }

We now ask how a more worker-friendly bargaining protocol affects aggregate stakeholder value. 
The objective is the number of filled jobs, $(N-v)$, times the equilibrium per-match stakeholder surplus $S_\omega + S_f$.
Throughout this section, ``stakeholder value'' refers to the sum of parties' continuation surpluses relative to outside states $\bp{(W-U)+(J-V)}$, not to utilitarian welfare; bargaining affects this object through equilibrium outside values and (under free entry) through tightness and entry.\footnote{%
In the canonical DMP model with fixed tightness and fixed employment levels, a standard flow welfare measure $\mathcal{W}=(1-u)y+ub-\kappa v$ is invariant to the wage split because wages are transfers.
Our criterion instead aggregates private continuation surpluses, which is the natural governance object when ``stakeholders'' are defined as workers and firms and the outside values $(U,V)$ matter for bargaining and participation.
We provide a brief comparison with flow welfare in Appendix~\ref{a:flow_welfare}.}
This criterion is distribution-neutral within matches: it adds workers' and firms' surpluses without assigning different weights.

At the bargaining stage, total match surplus is pinned down by output and outside values, so changing worker bargaining power only reallocates surplus within the match. 
The reason ``governance'' matters here is that equilibrium wages affect outside values $(U,V)$ and, under free entry, the number of operating firms and market tightness. 
When we refer below to ``per-match stakeholder surplus,'' we therefore mean the equilibrium object $S_\omega+S_f$ (with $(U,V,\theta)$ adjusting), not the bargaining-stage pie holding $(U,V)$ fixed.

\paragraph{Equilibrium with no free entry}
Consider first the case of no free entry, with a fixed number of firms $N$ and an associated fixed tightness $\theta$. 
Aggregate stakeholder value is
\[
    (N - v) \bp{S_\omega + S_f} = 
    \frac{\theta q(\theta)}{\delta + \theta q(\theta)}
    \left[
    \frac{y - w + \kappa}{1 - \beta(1 - \delta - q(\theta))} + \frac{w - b}{1 - \beta(1 - \delta - \theta q(\theta))}
    \right].
\]
Since the employment factor $\frac{\theta q(\theta)}{\delta+\theta q(\theta)}$ does not depend on $w$ when $\theta$ is fixed, the effect of a higher wage on stakeholder value is governed by the per-match derivative:
\[
    \frac{\partial (S_\omega+S_f)}{\partial w}
    =\frac{1}{1-\beta(1-\delta-\theta q(\theta))}
    -\frac{1}{1-\beta(1-\delta-q(\theta))}
    \gtrless 0
    \quad \Longleftrightarrow \quad
    \theta \lessgtr 1.
\]
Thus, when the market is slack ($\theta<1$), shifting value toward workers through a higher wage raises per-match stakeholder value. 
The economic reason is the asymmetry in continuation prospects: when $\theta<1$, unemployed workers find jobs more slowly than vacancies are filled ($p(\theta)=\theta q(\theta)<q(\theta)$), so a marginal increase in the current wage benefits workers more than it harms firms through outside options.

\paragraph{Free entry equilibrium}
Under free entry, changing bargaining power affects not only per-match surplus but also equilibrium tightness and the measure of operating firms. 
With identical firms, aggregate stakeholder value can be written as
\[
    (N-v)\bp{S_\omega+S_f}
    =\frac{\theta q(\theta)}{\delta+\theta q(\theta)}
    \bs{
    \frac{w-b}{1-\beta\bp{1-\delta-\theta q(\theta)}}
    \;+\;
    \frac{y-w+\kappa}{1-\beta\bp{1-\delta-q(\theta)}}
    },
\]
where $(w,\theta)$ solve the equilibrium conditions \eqref{eq:DMP_beveridge} and \eqref{eq:DMP_eqw_FE} (equivalently \eqref{eq:DMP_tightness_condition}).
While a closed-form optimum is not available, the comparative statics are transparent. 
Higher worker bargaining power $\phi$ raises the negotiated wage at a given tightness, shifting the wage-setting curve up. 
In equilibrium, this lowers tightness $\theta$; since $q(\theta)$ is decreasing, vacancy filling increases and job finding falls. 
The resulting increase in $q(\theta)$ partly mitigates the profitability loss from higher wages by shortening vacancy durations, but free entry also implies fewer firms operate in equilibrium, reducing employment. 
This employment effect is the central force pushing the stakeholder-value optimum toward low bargaining power.

\begin{figure}[t]
\caption{Optimal corporate governance in the DMP model}
\label{f:dmp_simulation}
\includegraphics[scale=0.38]{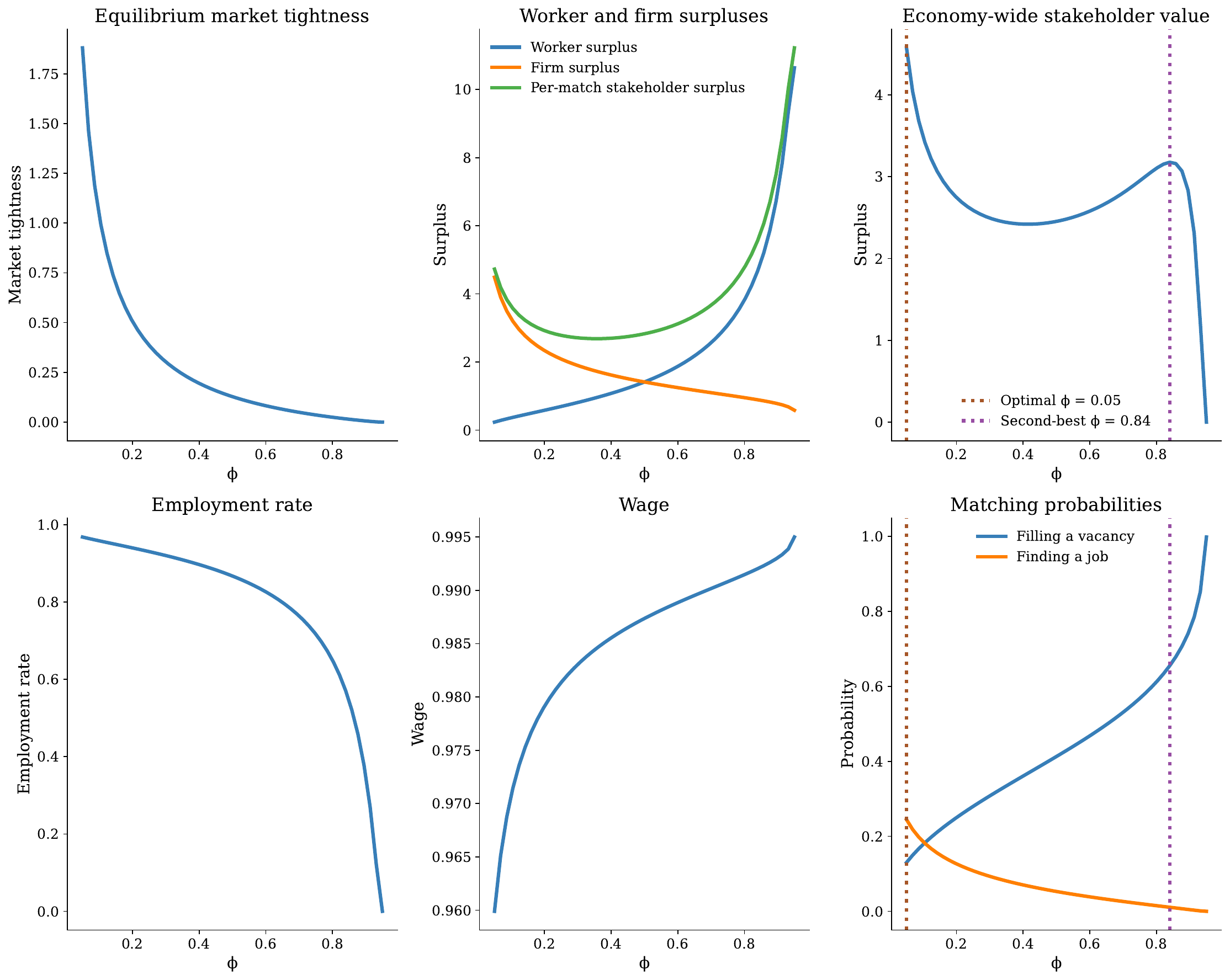}
\note[Note]{
The six panels depict the equilibrium in the DMP model, where responsible firms optimize corporate governance under free entry. 
Vertical lines mark the global maximizer of economy-wide stakeholder value and, when present, a distant local maximizer, reflecting the trade-off between per-match surplus and employment.
The model is calibrated following \cite{hagedorn2008cyclical} and \cite{hansel2024solving}, assuming a decreasing vacancy-filling probability $q(\theta) = (1 + \theta^\alpha)^{-1/\alpha}$ with $\alpha = 0.407$, and parameter values set as $y = 1$, $b = 0.9$, $\kappa = 0.584$, $\beta = (0.99)^{1/12}$, and $\delta = 0.0081$. 
The value of $\phi$ ranges between 0.05 and 0.95.
}
\end{figure}

\paragraph{Calibration and simulations}
Figure \ref{f:dmp_simulation} illustrates these forces under the calibration described in the caption. 
As $\phi$ rises, wages increase while equilibrium tightness falls, which raises vacancy filling and lowers job finding. 
Per-match stakeholder surplus varies non-monotonically with $\phi$, but aggregate stakeholder value is highest at very low bargaining power in our calibration because employment is then close to one (here at $\phi=0.05$).
Away from this extreme, a second (distant) local maximum can emerge at high $\phi$ (here around $\phi=0.84$), reflecting the trade-off between higher per-match surplus and lower employment.

\begin{figure}[t]
\caption{Optimal corporate governance with elastic labor supply}
\label{f:dmp_simulation_elastic_supply}
\includegraphics[scale=0.4]{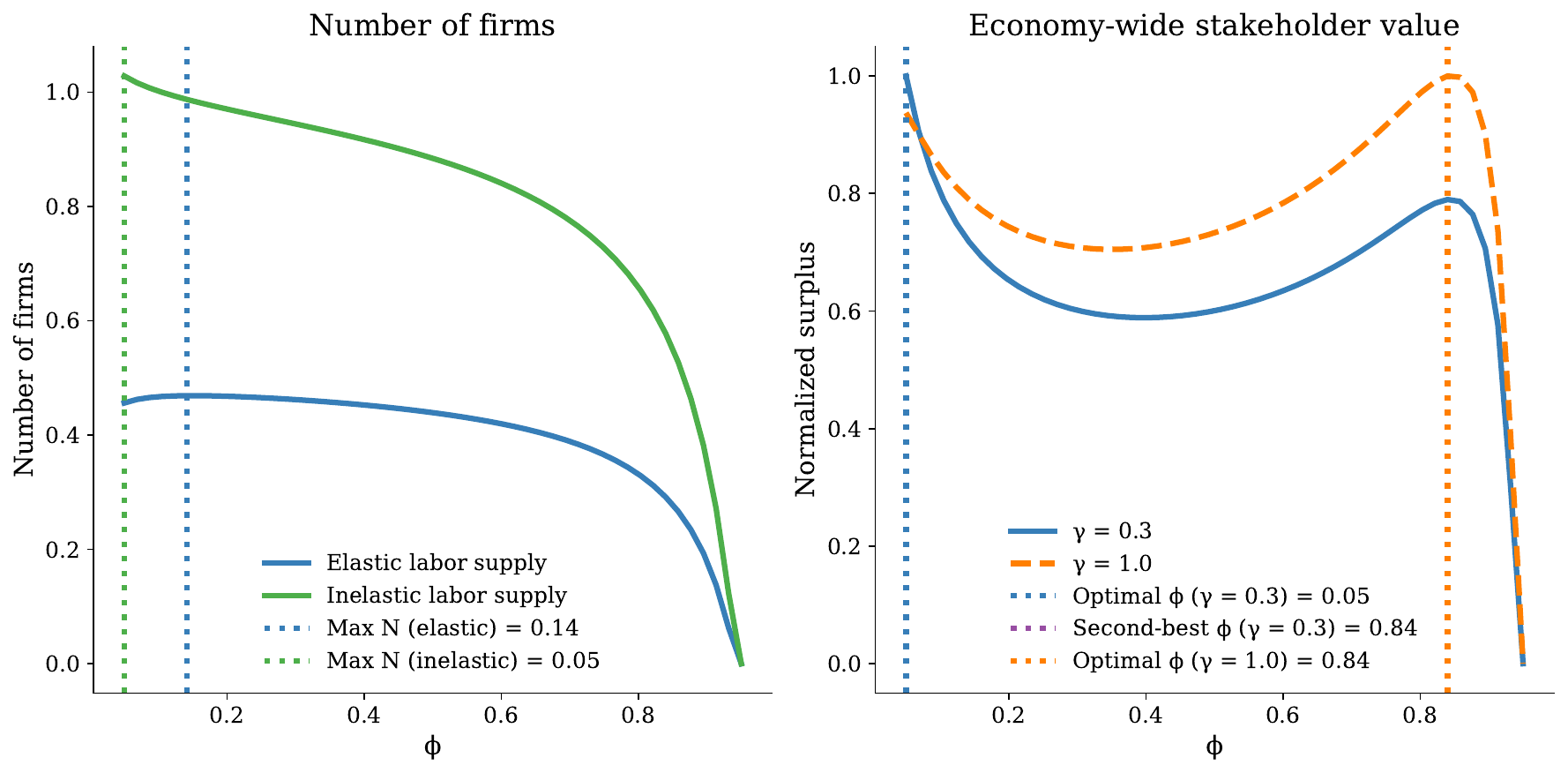}
\note[Note]{
The figure reports the equilibrium under free entry with elastic labor supply $L(w) = \left(\frac{w - b}{b}\right)^\gamma$.
The parametrization matches that of Figure \ref{f:dmp_simulation}.
The left panel reports the number of firms under elastic ($\gamma = 0.3$) and inelastic labor supply, with vertical lines marking the respective maximizers.
The right panel compares normalized economy-wide stakeholder value for two elasticities: $\gamma = 0.3$ (solid), the median extensive-margin estimate of \cite{blundell2013extensive}, and $\gamma = 1.0$ (dashed).
Vertical lines mark the global maximizer for each elasticity and, when present, a distant local maximizer.
Under low elasticity the global optimum remains at low $\phi$; under high elasticity the participation channel dominates and the high-$\phi$ peak becomes the global optimum.
}
\end{figure}

With elastic labor supply, a higher wage also raises participation, which attenuates the employment decline (Figure \ref{f:dmp_simulation_elastic_supply}). 
In our calibration the global maximizer of economy-wide stakeholder value remains at low $\phi$, but the high-$\phi$ local maximum becomes more pronounced, providing a second-best rationale for stronger worker bargaining power when bargaining power cannot be pushed close to zero.
The ranking of the two local maxima depends on the extensive-margin elasticity $\gamma$ (right panel of Figure \ref{f:dmp_simulation_elastic_supply}).
Because wages are compressed near $b$ under this calibration, the participation response $L(w)$ must grow fast enough to offset the decline in the employment rate as $\phi$ rises.
For low elasticities (including the baseline $\gamma=0.3$), the employment effect dominates and the global optimum remains at low $\phi$.
For sufficiently high elasticities ($\gamma \gtrsim 0.85$ under our parameterization), the participation channel dominates and the high-$\phi$ local maximum becomes the global optimum, providing a first-best rather than second-best rationale for strong worker bargaining power.
The critical threshold is lower when outside options are weak relative to productivity (low $b/y$), since the wage gap between the two optima is then wider.

\subsection{A subset of firms giving more power to the workers}
We now turn to a positive exercise: only a subset of firms commits to a more worker-friendly bargaining protocol. 
Specifically, a fraction $\mu$ of firms raises worker bargaining power from $\phi$ to $\phi_{\Rc}>\phi$.
As above, responsibility is modeled as a commitment to a bargaining protocol (a higher worker share of match surplus), which is the natural within-match policy lever in the DMP environment.
The proofs of the propositions are provided in Appendix \ref{a:add_computations_DMP}.

\paragraph{Setup}
Consider a DMP-style labor market where firms and workers engage in wage bargaining.
A fraction $\mu$ of “responsible firms” adopt a higher bargaining power parameter for workers, $\phi_{\Rc}$, while the remaining $1-\mu$ firms retain the original parameter $\phi$.
This change does not alter the total surplus at the firm level but may influence the overall market equilibrium.
For simplicity, we assume that matching probabilities depend only on aggregate tightness $\theta$ and not on wages or firm type, so responsible firms face the same vacancy-filling probability $q(\theta)$ and pay the same vacancy cost $\kappa$ as regular firms.
The wage equations for the two types of firms are:
\begin{equation}\label{eq:DMP_rf_regfirm}
    w = \phi \bp{y - V (1 - \beta )} + U \bp{1 - \phi} (1 - \beta) 
\end{equation}
for the regular firm and
\begin{equation}\label{eq:DMP_rf_respfirm}
    w_{\Rc} = \phi_{\Rc} \bp{y - V_{\Rc} (1 - \beta )} + U \bp{1 - \phi_{\Rc}} (1 - \beta) 
\end{equation}
for the responsible one.

\paragraph{Two-type value system}
Let $W$ and $W_{\Rc}$ denote the values of being employed at a regular and responsible firm, respectively, and let $U$ be the value of unemployment.
Unemployed workers meet vacancies at the aggregate job-finding probability $p(\theta)=\theta q(\theta)$; conditional on meeting a vacancy, the probability the firm is responsible equals the vacancy share, which under our assumptions coincides with the firm share $\mu$.\footnote{%
Since all unmatched firms post vacancies and all face the same separation probability $\delta$ and filling probability $q(\theta)$, each firm spends the same fraction of time vacant, so vacancy shares mirror firm shares.}
The stationary Bellman system is:
\[
    U=b+\beta\bs{(1-p(\theta))U+p(\theta)\bs{(1-\mu)W+\mu W_{\Rc}}},
\]
\[
    W=w+\beta\bs{(1-\delta)W+\delta U},\qquad
    W_{\Rc}=w_{\Rc}+\beta\bs{(1-\delta)W_{\Rc}+\delta U},
    \]
\[
    V=-\kappa+\beta\bs{(1-q(\theta))V+q(\theta)J},\qquad
    V_{\Rc}=-\kappa+\beta\bs{(1-q(\theta))V_{\Rc}+q(\theta)J_{\Rc}},
\]
\[
    J=y-w+\beta\bs{(1-\delta)J+\delta V},\qquad
    J_{\Rc}=y-w_{\Rc}+\beta\bs{(1-\delta)J_{\Rc}+\delta V_{\Rc}}.
\]
Equations \eqref{eq:DMP_rf_regfirm}--\eqref{eq:DMP_rf_respfirm} are the Nash-bargaining wages associated with these continuation values.

\begin{remark}\label{rem:no_rf_under_fe}
Under the maintained assumptions (common $\kappa$, common $q(\theta)$, difference only in $\phi_{\Rc}>\phi$), responsible firms have strictly negative vacancy value under free entry and do not enter.
This is immediate: higher wages with identical costs violate the zero-profit condition.
\end{remark}
\noindent This result reflects the simplifying assumption that responsibility operates only through bargaining power and not through the retention or recruiting channels identified in Section~\ref{sec:micro_model}; endogenizing those margins could sustain coexistence under free entry.

\paragraph{Equilibrium with no free entry}
Given Remark \ref{rem:no_rf_under_fe}, we focus on a scenario without free entry, where market tightness $\theta$ is fixed by the number of firms. 
This allows for the coexistence of both types of firms.

In this setting, we start by establishing the following proposition regarding the equilibrium wages of a responsible firm.
\begin{proposition}\label{prop:DMP_rf_wages}
    In an equilibrium with fixed market tightness $\theta < 1$, responsible firms pay higher wages than ordinary firms, provided that match surplus is positive at zero wages, i.e.\ $(y+\kappa)/\bigl(1-\beta(1-\delta-q(\theta))\bigr)>b/\bigl(1-\beta(1-\delta-\theta q(\theta))\bigr)$.
\end{proposition}
The condition requires that the match is productive even at $w=0$, i.e., that the total continuation value created by a match exceeds the worker's outside value.

\begin{proposition}\label{prop:DMP_w0}
    The introduction of responsible firms in a market with no free entry raises the wages in both responsible and regular firms above the initial wage level.
\end{proposition}
Responsible firms raise the unemployment value $U$ through the prospect of receiving a higher wage, strengthening workers' outside option in bargaining with \emph{both} types and shifting up regular-firm wages.

\begin{proposition}\label{prop:DMP_wR_monotonicity}
    In a stationary equilibrium with no free entry, the wage gap between responsible and ordinary firms decreases monotonically as the fraction $\mu$ of responsible firms increases.
\end{proposition}
As $\mu$ increases, unemployed workers are more likely to match with a high-wage responsible firm, which raises the value of unemployment $U$.
This strengthens workers' outside option in bargaining with \emph{both} firm types, so both $w$ and $w_{\Rc}$ typically rise with $\mu$.
The gap $w_{\Rc}-w$ nevertheless shrinks because ordinary firms’ wage responds more strongly to the induced increase in $U$.

\subsection{Discussion}
The DMP environment differs from the wage-posting framework of Section~\ref{sec:BM} in two key respects: wages are pinned down by bargaining rather than posted, and vacancy creation links wages to market tightness and employment.
These features make corporate governance---modeled here as the worker bargaining weight $\phi$---a lever that reshapes not only the within-match split but also equilibrium outside values, tightness, and the mass of operating firms.

\paragraph{Governance and stakeholder value}
The central normative result is that the effect of raising worker bargaining power on per-match stakeholder value hinges on a single observable: market tightness.
When the market is slack ($\theta<1$), unemployed workers find jobs more slowly than vacancies are filled ($p(\theta)<q(\theta)$), so a marginal dollar transferred from the firm to the worker raises the worker's continuation value by more than it lowers the firm's.
Higher wages therefore raise per-match stakeholder surplus in slack markets and lower it in tight ones.
Under free entry, however, higher bargaining power also reduces equilibrium tightness and the number of operating firms, creating a sharp employment--surplus trade-off.
In our calibration, the employment effect dominates and economy-wide stakeholder value is maximized at low $\phi$, but a second local maximum emerges at high $\phi$ where per-match surplus is large.
Elastic labor supply can reverse the ranking: when the extensive-margin elasticity is high enough ($\gamma\gtrsim 0.85$ in our parameterization), the participation response offsets the employment decline and the high-$\phi$ peak becomes the global optimum.
This suggests that the case for strong worker bargaining power is strongest in markets where labor supply is responsive to wages and outside options are weak relative to productivity.

\paragraph{Spillovers and wage convergence}
The positive results on two-type coexistence reveal an equilibrium spillover channel absent from the firm-level and wage-posting analyses.
When a fraction $\mu$ of firms commits to higher bargaining power, unemployed workers face a chance $\mu$ of matching with a high-wage firm in each meeting, which raises the value of unemployment $U$.
Since $U$ enters the bargaining problem of \emph{both} firm types as the worker's outside option, wages at ordinary firms rise even though those firms' bargaining protocol has not changed.
The wage gap $w_{\Rc}-w$ nevertheless shrinks as $\mu$ increases, because the coefficient linking ordinary-firm wages to $U$ is larger: ordinary firms' wages start further from the outside option and therefore respond more elastically to improvements in it.
In the limit $\mu\to 1$, all firms are responsible and the gap vanishes, but along the transition every additional responsible firm benefits workers at ordinary firms.

\paragraph{Coexistence and model limitations}
Under our maintained assumptions, responsible firms cannot coexist with profit-maximizers under free entry (Remark~\ref{rem:no_rf_under_fe}).
This is a mechanical consequence of the modeling choice: if two firm types differ only in the wage they pay, free entry selects the low-wage type.
We therefore analyze the two-type economy under limited competition, where tightness is pinned down by a fixed mass of firms.
This limitation points to a natural extension.
The micro-level analysis of Section~\ref{sec:micro_model} shows that responsibility reduces turnover and improves recruiting through higher worker surplus, which maps into lower effective vacancy costs or higher effective filling rates.
Endogenizing these channels---for example by letting $\kappa$ or $q$ depend on the firm's wage policy---could offset the higher wage bill and sustain responsible firms under free entry.
Exploring this direction would connect the firm-level advantages documented in Section~\ref{sec:micro_model} with the equilibrium tightness mechanism of the DMP framework.

\paragraph{Relation to the other sections}
The three environments studied in the paper illuminate different facets of responsible wage setting.
The firm-level model (Section~\ref{sec:micro_model}) identifies a wage-cost wedge: responsibility lowers the perceived marginal cost of wages, generating higher wages and more generous employment policies, with the wedge largest when outside options are weak and jobs are persistent.
The wage-posting model (Section~\ref{sec:BM}) translates this wedge into an equilibrium wage distribution, showing that responsible firms sort to the upper tail and can generate a segmented high-wage sector.
The DMP model adds two dimensions that those environments abstract from: endogenous market tightness, which links wages to aggregate employment through entry or the mass of operating firms, and bargaining, which transmits responsible governance into workers' outside options and hence into wages at all firms.
The governance and spillover results are therefore specific to the DMP setting.%, while the survival limitation reflects its deliberate simplification of the retention and recruiting margins that the micro model endogenizes.

\section{Conclusions}\label{sec:conclusions}

This paper develops a theory of responsible employers in frictional labor markets, where firms have wage-setting power because workers face search frictions and imperfect mobility.
We model responsibility as valuing worker welfare alongside profits, and we study how this objective reshapes wage setting, vacancy creation, and equilibrium outcomes across three canonical environments.

At the firm level (Section~\ref{sec:micro_model}), responsibility operates as a wedge in the firm’s perceived wage cost: raising wages is privately more attractive because it directly raises the welfare term the firm internalizes.
This wedge implies a more generous employment policy than a profit-maximizing firm would choose, but its magnitude is state-dependent.
When outside options are weak and jobs are persistent, the welfare gain from a higher wage is large; when outside options are strong and turnover is high, the marginal welfare gains from additional wage increases are smaller and the responsible firm behaves closer to a standard monopsonist.

At the market level, the wage-posting model (Section~\ref{sec:BM}) shows how responsibility leaves a clear footprint on the wage distribution.
Responsible firms are pulled toward higher wages, which can generate wage segmentation and sorting of firm types across a low-wage and a high-wage segment.
The direction and strength of sorting depend on productivity differences and on the efficiency-wage channel, and in the limit of full responsibility responsible firms bunch at the highest sustainable wage.

Finally, the DMP model (Section~\ref{sec:DMP}) highlights how responsibility interacts with market tightness and entry through the bargaining protocol that pins down wages.
Normatively, increasing worker bargaining power can raise per-match stakeholder value in slack markets, but under free entry it also reduces profitability and equilibrium employment by shrinking the mass of operating firms; with elastic labor supply, participation responses partly mitigate this trade-off.
Under limited competition, responsible firms can coexist with ordinary firms, pay higher wages in slack markets, and raise ordinary-firm wages by improving workers' outside option.

More broadly, the analysis clarifies that ``responsible wage setting'' is not a single rule such as always paying above-market wages.
In frictional labor markets, its implications depend on outside options, turnover, and equilibrium feedback through tightness.
An important direction for future work is to incorporate additional margins that shape responsibility and market structure---such as endogenous entry costs that differ across firm types, other stakeholder groups, and policies that affect mobility frictions and bargaining institutions.

\bibliography{Feb02}

\clearpage

%%%%%%%%%%%%%%%%%%%%%%%%%%%%%%%%%%%%%%%%%%%%%%%%%%%%%%%%%%%%%%%%%%%%%%%%%%%%%%%%
%% APPENDIX

\appendix

\section{Notation}\label{a:notation}
This appendix provides a compact guide to symbols that are reused across the three environments.
%The final column highlights the main notational collisions (same letter, different meaning) to help cross-reading.
\begin{table}[h!]
\caption{Notation and symbol reuse across environments}
\scriptsize
\setlength{\tabcolsep}{3pt}
\renewcommand{\arraystretch}{1.0}
\begin{tabular*}{\textwidth}[]{@{}p{1.4cm}@{\extracolsep\fill}p{2.2cm}p{2.2cm}p{2.2cm}p{5.8cm}@{}}
	\toprule
	Symbol
	& Sec.~\ref{sec:micro_model} &  Sec.~\ref{sec:BM} &  Sec.~\ref{sec:DMP}  &  Notes / collisions \\
	\midrule
$\beta$ & discount factor & --- & discount factor & Sec.~\ref{sec:BM} uses rates $(\lambda,\delta)$ instead. \\
$\lambda,\delta$ & --- & offer/sep.\ rates & --- & Continuous-time rates; $\delta$ reused as sep.\ prob.\ in Sec.~\ref{sec:micro_model}--\ref{sec:DMP}. \\
$\eta$ & resp.\ weight & resp.\ weight & --- & Sec.~\ref{sec:DMP} uses $\phi$ instead. Not $\eta_c$ (effort cost in Sec.~\ref{sec:BM} calibration). \\
$\phi,\,\phi_{\Rc}$ & --- & --- & bargaining power & Primitive; $\phi_{\Rc}>\phi$ for responsible firms. \\
$\varphi(\theta)$ & --- & --- & wage weight & Derived from $\phi$ and $\theta$; not a primitive. \\
	\addlinespace[0.2em]
$u_k,\,u$ & unemp.\ mass & unemp.\ rate & unemployment & Stock or rate; not utility. \\
$U_k,\,U$ & worker value & --- & unemp.\ value & Bellman value in both sections. \\
$W_k$ & worker value & --- & $W,\,W_{\Rc}$ & Sec.~\ref{sec:DMP} distinguishes regular/responsible matches. \\
$S_k$ & worker surplus (PV) & --- & --- & PV relative to $U_k$; flow via $(1-\beta)S_k$. \\
$S(\cdot)$ & --- & flow surplus & --- & $S(w)=\tilde u(w)-b$; flow normalization. \\
$S_\omega,\,S_f$ & --- & --- & worker/firm surplus & Bargaining surpluses $W\!-\!U$ and $J\!-\!V$. \\
$v_k,\,V$ & vacancies; --- & --- & vacancies; vac.\ value & $v_k$ choice; $V$ Bellman value. \\
$J$ & --- & --- & job value & Bellman value of a filled job. \\
$w_k,\,w$ & posted wage & posted wage & bargained wage & Choice in Sec.~\ref{sec:micro_model}--\ref{sec:BM}; bargaining in Sec.~\ref{sec:DMP}. \\
$y,\,b$ & output; outside opt. & output; outside opt. & output; outside opt. & Same primitives across sections. \\
	\addlinespace[0.2em]
$\theta_k,\,\theta$ & $\bar v_k/u_k$ & --- & $v/u$ & Firm-level in Sec.~\ref{sec:micro_model}; aggregate in Sec.~\ref{sec:DMP}. \\
$\hat q_k,\,\bar q_k$ & meeting/fill prob. & --- & --- & Vacancy filling; decreasing in tightness. \\
$q(\theta)$ & --- & --- & fill prob. & $(1+\theta^\alpha)^{-1/\alpha}$; decreasing in $\theta$. \\
$p(\theta)$ & --- & --- & job-finding prob. & $\theta\,q(\theta)$; increasing in $\theta$. \\
$\bar p_k^{u},\,\bar p_k^{e}$ & offer arrivals & --- & --- & Unemp.\ vs on-the-job (reduced-form). \\
$p_k(w)$ & improving-offer & --- & --- & $(1\!-\!F_k(w))\bar p_k^{e}$; not $p(\theta)$. \\
$F_k,\,F$ & offer CDF & wage CDF(s) & --- & Offer vs equilibrium distributions. \\
	\addlinespace[0.2em]
$\gamma_k$ & sep.\ prob.\ (endog.) & --- & --- & $\gamma_k(w_k)$; decreasing in wage. \\
$\delta$ & sep.\ prob.\ (exog.) & sep.\ rate & sep.\ prob.\ (exog.) & Exogenous in Sec.~\ref{sec:micro_model},\ref{sec:DMP}; Poisson rate in Sec.~\ref{sec:BM}. \\
$\kappa$ & --- & --- & vacancy cost & Per-period cost of posting a vacancy. \\
$N$ & --- & --- & mass of firms & Limited vs free entry. \\
$\alpha$ & --- & type-1 firm share & matching curvature & Firm share in Sec.~\ref{sec:BM} vs matching param.\ in Sec.~\ref{sec:DMP}. \\
$\mu$ & --- & --- & resp.\ firm share & Fraction with $\phi_{\Rc}$ in two-type exercise. \\
\bottomrule
\end{tabular*}
\note[Note]{The ``Notes'' column flags the main reuse/collision risks. The paper defines all objects locally; this table highlights the handful of symbols whose meaning changes across environments.}
\label{tab:notation}
\end{table}

\section{Additional computations and proofs for the firm-level model}\label{a:add_computations_micro}

\subsection{Derivation of the profit function}
This subsection makes explicit the expectation operator underlying the firm's per-period profit function \eqref{eq:profit}.
To make the expectation explicit, let $\psi(\ell,v,w)$ denote the joint density of employment outcomes $\ell\in[0,v]$ conditional on posted $(v,w)$, and assume independence across segments so that $\psi(\ell,v,w)=\prod_{k=1}^m \psi_k(\ell_k,v_k,w_k)$.
Then
\[
\E{f\of{\ell,w}}=\int_0^v \psi(\ell,v,w) f\of{\ell,w} d\ell.
\]
Moreover, given posted $(v,w)$, $\E{\ell_k}=\int_0^v \psi(\ell,v,w)\,\ell_k\,d\ell$, and therefore the expected wage bill decomposes as
\[
\int_0^v \psi(\ell,v,w)\,w\ell\,d\ell=\sum_{k=1}^m w_k\,\E{\ell_k}.
\]
Finally, in the main text we summarize hiring in each segment $k$ by the expected number of filled jobs,
\[
\E{\ell_k}=v_k\,\bar q_k(\theta_k,w_k),\qquad \theta_k:=\frac{v_k}{u_k},\qquad
\bar q_k(\theta_k,w_k):=\frac{\hat q_k(\theta_k,w_k)}{\hat q_k(\theta_k,w_k)+\gamma_k(w_k)}.
\]
These definitions rationalize the compact representation \eqref{eq:profit} used in the main text.

\subsection{Responsible firm's first-order conditions}

We derive the first-order conditions for the responsible firm's problem stated in Section \ref{sec:micro_model}.
The objective can be written as
\[
\mathcal{J} = \frac{1}{1-\beta}\Bigl(\E{f\of{\ell,w}}-\sum_{j=1}^m w_j \E{\ell_j}+\eta\sum_{j=1}^m (1-\beta)S_j(w_j)\E{\ell_j}\Bigr).
\]

\paragraph{Vacancies}
We differentiate the objective $\mathcal{J}$ with respect to vacancies in market $k$, $v_k$:
\[
\pd{\mathcal{J}}{v_k} = \frac{1}{1-\beta}\pd{}{v_k}\E{f\of{\ell,w}} - \frac{w_k}{1-\beta}\pd{\E{\ell_k}}{v_k} + \eta S_k(w_k)\pd{\E{\ell_k}}{v_k} = 0.
\]
Using $\E{\ell_k}=v_k\,\bar q_k(\theta_k,w_k)$ and $\theta_k=v_k/u_k$, we have
\[
\pd{\E{\ell_k}}{v_k}=\bar q_k(\theta_k,w_k)+\theta_k\,\pd{\bar q_k(\theta_k,w_k)}{\theta_k},
\]
and we maintain the regularity condition $\bar q_k+\theta_k\,\pd{\bar q_k}{\theta_k}>0$ (equivalently $\varepsilon_{\bar q_k:\theta_k}<1$) so that expected employment increases with vacancies.
Define marginal revenue from vacancies as
\[
MR(v_k):=\pd{}{v_k}\E{f\of{\ell,w}}.
\]
Using $\E{f\of{\ell,w}}=\int_0^v \psi(\ell,v,w) f\of{\ell,w}d\ell$, an explicit integral representation is:
\begin{equation*}
    \begin{split}
        MR(v_k)
        &= \int_0^{v_{-k}} \psi(v_k, \ell_{-k}, v, w) f\of{v_k, \ell_{-k}, w} d \ell_{-k} + \int_0^v \psi_{v_k}(\ell, v, w) f\of{\ell, w} d\ell.
    \end{split}
\end{equation*}

Substituting into the first-order condition yields
\[
MR(v_k)=\bs{w_k-\eta(1-\beta)S_k(w_k)}\pd{\E{\ell_k}}{v_k},
\]
which is equivalent to \eqref{eq:FOC_v_REF}.

\paragraph{Wage}
We differentiate the objective $\mathcal{J}$ with respect to the wage in market $k$, $w_k$. 
This involves differentiating the expected revenue, the expected wage cost, and the expected worker surplus.

Define marginal revenue from wages as $MR(w_k):=\pd{}{w_k}\E{f\of{\ell,w}}$.
Using $\E{f\of{\ell,w}}=\int_0^v \psi(\ell,v,w) f\of{\ell,w}d\ell$, we can write:
\begin{equation*}
    \begin{split}
        MR(w_k)
        = \int_0^v \left[ \psi_{w_k}(\ell, v, w)f\of{\ell, w} + \psi(\ell, v, w) f_{w_k}(\ell, w) \right] d\ell.
    \end{split}
\end{equation*}
Under the efficiency-wage production structure $f(\ell,w)=f(\ell_1 e_1(w_1),\dots,\ell_m e_m(w_m))$, we have $f_{w_k}(\ell,w)=f_k(\tilde\ell)\,\ell_k\,e_k'(w_k)$, where $\tilde\ell:=\bp{\ell_1 e_1(w_1),\dots,\ell_m e_m(w_m)}$.

The derivative of the wage cost is:
\[ \frac{\partial}{\partial w_k} \left[ w_k \E{\ell_k} \right] = \E{\ell_k} + w_k \frac{\partial \E{\ell_k}}{\partial w_k}
\]

The derivative of the responsibility term is:
\[ 
\eta\frac{\partial}{\partial w_k} \left[ S_k(w_k) \E{\ell_k} \right] = \eta\frac{\partial S_k}{\partial w_k} \E{\ell_k} + \eta S_k(w_k) \frac{\partial \E{\ell_k}}{\partial w_k}
\]

Setting the full derivative $\frac{\partial \mathcal{J}}{\partial w_k} = 0$:
\[
MR(w_k) = \left( \E{\ell_k} + w_k \pd{\E{\ell_k}}{w_k} \right) - \eta(1-\beta)\left( \pd{S_k}{w_k}\E{\ell_k} + S_k \pd{\E{\ell_k}}{w_k} \right).
\]

We group terms on the right-hand side by $\E{\ell_k}$ and $\pd{\E{\ell_k}}{w_k}$:
\[
MR(w_k) = \left[w_k - \eta(1-\beta)S_k\right]\pd{\E{\ell_k}}{w_k} + \left[1 - \eta(1-\beta)\pd{S_k}{w_k}\right]\E{\ell_k}
\]

To express this in terms of elasticities, we divide the entire equation by $\pd{\E{\ell_k}}{w_k}$:
\[
\frac{MR(w_k)}{\pd{\E{\ell_k}}{w_k}} = \left[w_k - \eta(1-\beta)S_k\right] + \left[1 - \eta(1-\beta)\pd{S_k}{w_k}\right]\frac{\E{\ell_k}}{\pd{\E{\ell_k}}{w_k}}
\]

Now, we introduce the elasticities:
\[ \varepsilon_{\E\ell_k:w_k} = \frac{\partial\E{\ell_k}}{\partial w_k} \frac{w_k}{\E{\ell_k}} \implies \frac{\E{\ell_k}}{\pd{\E{\ell_k}}{w_k}} = \frac{w_k}{\varepsilon_{\E\ell_k:w_k}} \]
and 
\[ \varepsilon_{S_k:w_k} = \frac{\partial S_k}{\partial w_k} \frac{w_k}{S_k} \implies \pd{S_k}{w_k} = \frac{S_k \varepsilon_{S_k:w_k}}{w_k} \]

Substituting these into the right-hand side:
\begin{align*} 
RHS &= \left[w_k - \eta(1-\beta)S_k\right] + \left[1 - \eta(1-\beta)\frac{S_k \varepsilon_{S_k:w_k}}{w_k}\right]\frac{w_k}{\varepsilon_{\E\ell_k:w_k}} \\ 
&= w_k - \eta(1-\beta)S_k + \frac{w_k}{\varepsilon_{\E\ell_k:w_k}} - \eta(1-\beta)S_k \frac{\varepsilon_{S_k:w_k}}{\varepsilon_{\E\ell_k:w_k}} \\ 
&= \left(w_k + \frac{w_k}{\varepsilon_{\E\ell_k:w_k}}\right) - \left(\eta(1-\beta)S_k + \eta(1-\beta)S_k \frac{\varepsilon_{S_k:w_k}}{\varepsilon_{\E\ell_k:w_k}}\right) \\
&= w_k\left(1 + \frac{1}{\varepsilon_{\E\ell_k:w_k}}\right) - \eta(1-\beta)S_k\left(1 + \frac{\varepsilon_{S_k:w_k}}{\varepsilon_{\E\ell_k:w_k}}\right) 
\end{align*}
This yields the final expression for the wage FOC \eqref{eq:FOC_w_REF}.

\subsection{Proof of Lemma \ref{lemma:W_in_w}}
Fix segment $k$ and write $\tilde W_k(w):=\E[s>w]{W_k(s)}$.
Assume that $F_k$ admits a density $f_k$, so that $p_k(w)=\bar p_k^{e}(1-F_k(w))$ is differentiable and
\[
p_k(w)\tilde W_k(w)=\bar p_k^{e}\int_w^{+\infty}W_k(s)\,dF_k(s)
\quad\Rightarrow\quad
\frac{d}{dw}\Bigl[p_k(w)\tilde W_k(w)\Bigr]=p_k'(w)\,W_k(w),
\]
where $p_k'(w)=-\bar p_k^{e} f_k(w)$.

For any fixed effort level $e$, solving \eqref{eq:worker_surplus} for the value of employment yields
\[
\bar W_k(w;e)
=\frac{w-c_k(e)+\beta\Bigl[\gamma_k(e)U_k+\bigl(1-\gamma_k(e)\bigr)p_k(w)\tilde W_k(w)\Bigr]}
{1-\beta\bigl(1-\gamma_k(e)\bigr)\bigl(1-p_k(w)\bigr)}.
\]
At the worker's optimum, $W_k(w)=\bar W_k(w;e_k(w))$.
For brevity, when evaluating at the optimum we write $\bar W_k(w):=\bar W_k(w;e_k(w))$ and use envelope derivatives.
Since effort maximizes $W_k(w)$ (equivalently $S_k(w)=W_k(w)-U_k$), the envelope theorem gives
$W_k'(w)=\partial_w \bar W_k(w;e)\vert_{e=e_k(w)}$.
	Let $\mathcal D_k(w,e):=1-\beta\bigl(1-\gamma_k(e)\bigr)\bigl(1-p_k(w)\bigr)$.
	Using the quotient rule together with $\frac{d}{dw}\bigl[p_k(w)\tilde W_k(w)\bigr]=p_k'(w)W_k(w)$, the $p_k'(w)$ terms cancel and
	\[
	W_k'(w)=\frac{1}{\mathcal D_k(w,e_k(w))}=\frac{1}{1-\beta\bigl(1-\gamma_k(e_k(w))\bigr)\bigl(1-p_k(w)\bigr)}>0.
	\]
The inequality holds because $\beta<1$ and $\bigl(1-\gamma_k(e_k(w))\bigr)\bigl(1-p_k(w)\bigr)\in[0,1]$.

\subsection{Proof of Lemma \ref{lemma:e_in_w}}
Fix $k$ and consider the worker's problem of choosing effort $e_k$ to maximize $W_k(w)$.
The first-order condition (using $\gamma_k(w)=\gamma_k(e_k(w))$) can be written as
\[
c_k'(e_k)
=\beta\gamma_k'(e_k)\Bigl[U_k-\bigl(1-p_k(w)\bigr)W_k(w)-p_k(w)\tilde W_k(w)\Bigr],
\]
where $\tilde W_k(w)=\E[s>w]{W_k(s)}$ as above.
Since $\gamma_k'(e_k)<0$, this is equivalent to
\[
\frac{c_k'(e_k)}{-\gamma_k'(e_k)}
=\beta\Bigl[\bigl(1-p_k(w)\bigr)W_k(w)+p_k(w)\tilde W_k(w)-U_k\Bigr]
=:H_k(w).
\]

By convexity of $c_k$ and convexity of $\gamma_k$ (with $\gamma_k'<0$), the left-hand side is increasing in effort:
\[
\pd{}{e}\left(\frac{c_k'(e)}{-\gamma_k'(e)}\right)
=\frac{c_k''(e)\bigl(-\gamma_k'(e)\bigr)+c_k'(e)\gamma_k''(e)}{\bigl(-\gamma_k'(e)\bigr)^2}\ge 0.
\]
Moreover,
\[
H_k'(w)=\beta\bigl(1-p_k(w)\bigr)W_k'(w)>0,
\]
	using $\frac{d}{dw}\bigl[p_k(w)\tilde W_k(w)\bigr]=p_k'(w)W_k(w)$ and Lemma \ref{lemma:W_in_w} (and noting that $p_k(w)<1$ when $\bar p_k^{e}\in(0,1)$).
Therefore, since the effort condition equates an increasing function of $e_k$ to an increasing function of $w$, the optimal effort $e_k(w)$ is increasing in $w$.

\subsection{Proof of Proposition \ref{prop:surplus_behavior}}

Fix $k$ and suppress the segment index.

\paragraph{Effect of $w$}
Since the unemployment value $U_k$ does not depend on the worker's current wage, we have
\[
\pd{S_k(w)}{w}=\pd{W_k(w)}{w}>0
\]
by Lemma \ref{lemma:W_in_w}.

\paragraph{Effect of $\gamma_k(w)$}
Consider a marginal increase in the separation probability at wage $w$, i.e.\ in $\gamma_k(w)$ holding $\gamma_k(\cdot)$ fixed at other wages.
This pointwise change affects neither $U_k$ (which aggregates over wage offers) nor $\tilde W_k(w):=\E[s>w]{W_k(s)}$.
Rewrite \eqref{eq:worker_surplus} as
\[
	\mathcal D_k(w)\,W_k(w)=w-c_k(e_k(w))+\beta\gamma_k(w)U_k+\beta\bigl(1-\gamma_k(w)\bigr)p_k(w)\tilde W_k(w),
	\]
	where $\mathcal D_k(w):=1-\beta\bigl(1-\gamma_k(w)\bigr)\bigl(1-p_k(w)\bigr)$.
	Differentiating with respect to $\gamma_k(w)$ yields
	\[
	\pd{W_k(w)}{\gamma_k(w)}
	=\frac{\beta}{\mathcal D_k(w)}\Bigl[U_k-\bigl(1-p_k(w)\bigr)W_k(w)-p_k(w)\tilde W_k(w)\Bigr]<0,
	\]
because Lemma \ref{lemma:W_in_w} implies $\tilde W_k(w)\ge W_k(w)$ and $W_k(w)\ge U_k$ for any accepted wage.
Hence $\pd{S_k(w)}{\gamma_k(w)}<0$.

\paragraph{Effect of $\bar{p}_k^{u}$}
Hold $\bar p_k^{e}$ fixed.
Using $W_k(w)=U_k+S_k(w)$ in the unemployment value equation yields
\[
(1-\beta)U_k=b+\beta \bar p_k^{u}\,\E{S_k(w)}.
\]
Substituting this identity into the employment value equation \eqref{eq:worker_surplus} delivers a surplus recursion of the form
\[
S_k(w)
=\frac{w-c_k\of{e_k(w)}-b-\beta \bar p_k^{u}\,\E{S_k(w)}
+\beta\bigl(1-\gamma_k(w)\bigr)p_k(w)\,\E[s>w]{S_k(s)}}
{1-\beta\bigl(1-\gamma_k(w)\bigr)\bigl(1-p_k(w)\bigr)}.
\]
The associated Bellman operator is a contraction and is decreasing in $\bar p_k^{u}$, hence its unique fixed point $S_k(w)$ is decreasing in $\bar p_k^{u}$.

\subsection{Proof of Proposition \ref{prop:REF_vacancies}}
Compare the profit-maximizing vacancy condition \eqref{eq:FOC_v} with the REF condition \eqref{eq:FOC_v_REF}:
\[
\pd{}{v_k}\E{f\of{\ell,w}}
=w_k\,\pd{\E{\ell_k}}{v_k},
\qquad
\pd{}{v_k}\E{f\of{\ell,w}}
=\bs{w_k-\eta(1-\beta)S_k(w_k)}\,\pd{\E{\ell_k}}{v_k}.
\]
The two equations are identical except that the wage term $w_k$ is replaced by the effective wage cost $w_k-\eta(1-\beta)S_k(w_k)$, which yields the claim.

\subsection{Proof of Proposition \ref{prop:REF_markdown}}
Let $\varepsilon\equiv \varepsilon_{\E\ell_k:w_k}$ and write $a:=\eta(1-\beta)$.
Start from the REF wage condition \eqref{eq:FOC_w_REF} and use $S_k(w_k)\,\varepsilon_{S_k:w_k}=w_k\,\pd{S_k(w_k)}{w_k}$ to rewrite its right-hand side as
\[
\bp{1+\frac{1}{\varepsilon}}w_k-aS_k(w_k)\bp{1+\frac{\varepsilon_{S_k:w_k}}{\varepsilon}}
=w_k+\frac{w_k}{\varepsilon}\Bigl[1-a\Bigl(\frac{S_k(w_k)}{w_k}\varepsilon+\pd{S_k(w_k)}{w_k}\Bigr)\Bigr].
\]
Define $D_k$ as in \eqref{eq:varXi} and let $\varXi:=1/D_k$. Then the preceding display becomes
\[
\bp{1+\frac{D_k}{\varepsilon}}w_k
=\bp{1+\frac{1}{\varepsilon\,\varXi}}w_k.
\]
Comparing with the profit-maximizing condition \eqref{eq:FOC_w_elast} shows that the REF behaves as if the labor-supply elasticity were multiplied by $\varXi$.
Finally, since $S_k(w_k)>0$, $\pd{S_k(w_k)}{w_k}>0$ (Proposition \ref{prop:surplus_behavior}), and $\varepsilon>0$, we have $D_k<1$; together with the maintained regularity condition $D_k>0$, this implies $\varXi>1$.

\subsection{Proof of Corollary \ref{cor:REF_wage_decomposition}}
Let $\varepsilon\equiv \varepsilon_{\E\ell_k:w_k}$ and write $a:=\eta(1-\beta)$.
Starting from \eqref{eq:FOC_w_REF}, use $S_k(w_k)\,\varepsilon_{S_k:w_k}=w_k\,\pd{S_k(w_k)}{w_k}$ to rewrite the right-hand side as
\[
\bp{1+\frac{1}{\varepsilon}}w_k-aS_k(w_k)\bp{1+\frac{\varepsilon_{S_k:w_k}}{\varepsilon}}
=\bp{w_k-aS_k(w_k)}+\frac{1}{\varepsilon}\bp{w_k-a w_k\,\pd{S_k(w_k)}{w_k}}.
\]
Factoring out $w_k-aS_k(w_k)$ gives
\[
\bp{w_k-aS_k(w_k)}\bp{1+\frac{1}{\varepsilon\,\varGamma}},
\qquad
\varGamma:=\frac{1-a\frac{S_k(w_k)}{w_k}}{1-a\,\pd{S_k(w_k)}{w_k}},
\]
which yields the stated decomposition.

\subsection{Proof of Proposition \ref{prop:REF_Xi_behavior}}
Fix $k$ and write $\varepsilon\equiv \varepsilon_{\E \ell_k:w_k}$.
Recall from \eqref{eq:varXi} that $\varXi=1/D_k$, where
\[
D_k
=1-\eta(1-\beta)\left(\frac{S_k(w_k)}{w_k}\varepsilon+\pd{S_k(w_k)}{w_k}\right).
\]
For any scalar parameter $x$, $\pd{\varXi}{x}=-D_k^{-2}\pd{D_k}{x}$, so $\varXi$ decreases (resp.\ increases) in $x$ whenever $D_k$ increases (resp.\ decreases) in $x$.

Holding $\varepsilon$ fixed, Lemma \ref{lemma:stw} implies that $S_k(w_k)/w_k$ decreases with $\bar p_k^{u}$ and with $\gamma_k(w_k)$, and that $\pd{S_k(w_k)}{w_k}$ decreases with $\gamma_k(w_k)$.
Therefore $\pd{D_k}{\bar p_k^{u}}>0$ and $\pd{D_k}{\gamma_k(w_k)}>0$, which yields the first claim.
Finally, Lemma \ref{lemma:stw} implies that $\pd{S_k(w_k)}{w_k}$ increases with $w_k$ and that $S_k(w_k)/w_k$ increases with $w_k$ provided $\varepsilon_{S_k:w_k}>1$.
Under this maintained condition, $\pd{D_k}{w_k}<0$ and thus $\pd{\varXi}{w_k}>0$.

\subsection{Proof of Lemma \ref{lemma:stw}}
\paragraph{Surplus-to-wage ratio}
Fix $k$ and hold $w_k$ constant.
Since $w_k>0$, the ratio $S_k(w_k)/w_k$ inherits the monotonicity of $S_k(w_k)$.
Therefore it decreases with $\bar p_k^{u}$ and with $\gamma_k(w_k)$ by Proposition \ref{prop:surplus_behavior}.
Moreover,
\[
\pd{}{w_k}\left(\frac{S_k(w_k)}{w_k}\right)
=\frac{1}{w_k}\left(\pd{S_k(w_k)}{w_k}-\frac{S_k(w_k)}{w_k}\right),
\]
so $S_k(w_k)/w_k$ increases with $w_k$ provided $\varepsilon_{S_k:w_k}>1$.

\paragraph{Marginal surplus}
Since $U_k$ does not depend on the current wage, $\pd{S_k(w_k)}{w_k}=\pd{W_k(w_k)}{w_k}$.
Lemma \ref{lemma:W_in_w} gives the closed form
\[
\pd{S_k(w_k)}{w_k}
=\frac{1}{1-\beta\bigl(1-\gamma_k(w_k)\bigr)\bigl(1-p_k(w_k)\bigr)}.
\]
Holding $\gamma_k(w_k)$ fixed, $p_k(w_k)=\bar p_k^{e}\bigl(1-F_k(w_k)\bigr)$ implies
\[
\pd{}{\,\bar p_k^{e}}\pd{S_k(w_k)}{w_k}
=-\frac{\beta\bigl(1-\gamma_k(w_k)\bigr)\bigl(1-F_k(w_k)\bigr)}
{\left(1-\beta\bigl(1-\gamma_k(w_k)\bigr)\bigl(1-p_k(w_k)\bigr)\right)^2}<0.
\]
Similarly, holding $p_k(w_k)$ fixed,
\[
\pd{}{\,\gamma_k(w_k)}\pd{S_k(w_k)}{w_k}
=-\frac{\beta\bigl(1-p_k(w_k)\bigr)}
{\left(1-\beta\bigl(1-\gamma_k(w_k)\bigr)\bigl(1-p_k(w_k)\bigr)\right)^2}<0.
\]
Finally, $p_k'(w_k)=-\bar p_k^{e} f_k(w_k)<0$ and Lemma \ref{lemma:e_in_w} implies $\gamma_k'(w_k)<0$, so
\[
\pd{}{w_k}\pd{S_k(w_k)}{w_k}
=-\frac{\pd{}{w_k}\Bigl(1-\beta\bigl(1-\gamma_k(w_k)\bigr)\bigl(1-p_k(w_k)\bigr)\Bigr)}
{\left(1-\beta\bigl(1-\gamma_k(w_k)\bigr)\bigl(1-p_k(w_k)\bigr)\right)^2}>0.
\]

\subsection{Proof of Proposition \ref{prop:perfect_competition}}
Let $\varepsilon\equiv \varepsilon_{\E \ell_k:w_k}$.
For the profit-maximizing firm, \eqref{eq:FOC_w_elast} implies
\[
\frac{\pd{}{w_k}\E{f\of{\ell,w}}}{\pd{\E{\ell_k}}{w_k}}
=\bp{1+\frac{1}{\varepsilon}}w_k,
\]
so taking $\varepsilon\to\infty$ yields $\frac{\pd{}{w_k}\E{f\of{\ell,w}}}{\pd{\E{\ell_k}}{w_k}}\to w_k$.
For the responsible firm, \eqref{eq:FOC_w_REF} implies
\[
\frac{\pd{}{w_k}\E{f\of{\ell,w}}}{\pd{\E{\ell_k}}{w_k}}
=\bp{1+\frac{1}{\varepsilon}}w_k-\eta(1-\beta)S_k(w_k)\bp{1+\frac{\varepsilon_{S_k:w_k}}{\varepsilon}},
\]
so taking $\varepsilon\to\infty$ yields
$\frac{\pd{}{w_k}\E{f\of{\ell,w}}}{\pd{\E{\ell_k}}{w_k}}\to w_k-\eta(1-\beta)S_k(w_k)$.
This corresponds to the benchmark in which the wage is treated as effectively given at the relevant wage benchmark.

\section{Additional computations and proofs for the wage-posting model}\label{a:add_computations_BM}

\subsection{Proof of Proposition \ref{prop:BM_equal_prod}}
Assume, toward a contradiction, that a REF offers $w_2$ and a PMF offers $w_1$ with $w_2 < w_1$.
At the equilibrium, one must always have
\[ \ell(w_2) M_2(w_2) \ge \ell(w_1) M_2(w_1) \; \text{ and } \; \ell(w_1) Y_1(w_1) \ge \ell(w_2) Y_1(w_2)\]
Since $Y_2(w) = Y_1(w)$, the first inequality becomes
\[ 
\ell(w_2) \bs{Y_1(w_2) + \eta S(w_2)} 
 \ge 
\ell(w_1) \bs{Y_1(w_1) + \eta S(w_1)}
\]
Using the second inequality,
\[
\ell(w_2) \bs{Y_1(w_2) + \eta S(w_2)} 
 \ge
\ell(w_2) Y_1(w_2) + \ell(w_1) \eta S(w_1),
\]
implying $\ell(w_2)S(w_2) \ge \ell(w_1)S(w_1)$.
Since $w_2<w_1$ implies $\ell(w_2)<\ell(w_1)$ and $S(w_2)\le S(w_1)$ with $S(w_1)>0$ for accepted wages, the inequality cannot hold.

\subsection{Proof of Proposition \ref{prop:BM_REFs_mp}}
Assume, toward a contradiction, that a REF offers $w_2$ and a PMF offers $w_1$ with $w_2 < w_1$.
At the equilibrium, one must have
\[ 
\ell(w_2) M_2(w_2) \ge \ell(w_1) M_2(w_1) 
\text{ and } 
\ell(w_1) Y_1(w_1) \ge \ell(w_2) Y_1(w_2)
\]
Let $\Delta(w) = Y_2(w) - Y_1(w) = y_2(e^*(w)) - y_1(e^*(w)) > 0$.
Moreover, $\Delta$ is non-decreasing because
\[
\Delta'(w)=\bs{y_2'-y_1'}(e^*(w))\cdot e^{*\prime}(w)\ge 0
\]
by Assumption \ref{ass:ydiff} and $e^{*\prime}(w)>0$.
Rewrite the first inequality:
\[ 
\ell(w_2) \bs{Y_1(w_2) + \Delta(w_2) + \eta S(w_2)} 
 \ge 
\ell(w_1) \bs{Y_1(w_1) + \Delta(w_1) + \eta S(w_1)}
\]
Using the second inequality yields
\[
\ell(w_1) \bs{ Y_1(w_1) + \Delta(w_1) + \eta S(w_1)} 
 \ge 
\ell(w_2) Y_1(w_2) + \ell(w_1) \bs{ \Delta(w_1) + \eta S(w_1) }
\]
Combining gives
\[
\ell(w_2) \bs{\Delta(w_2) + \eta S(w_2)} 
 \ge 
\ell(w_1) \bs{\Delta(w_1) + \eta S(w_1)}
\]
Both $\ell$ and $S$ are increasing, $S(w)\ge 0$ for all accepted wages $w\ge w^r$, and $\Delta(w)> 0$ and is non-decreasing by Assumption \ref{ass:ydiff}.
Therefore, when $w_2<w_1$, we have simultaneously $\ell(w_2)< \ell(w_1)$ and $0<\Delta(w_2)+\eta S(w_2)\le \Delta(w_1)+\eta S(w_1)$, so the displayed inequality cannot hold. Contradiction.

% \subsection{Proof of Proposition \ref{prop:BM_ref_ge_pmf}}
% Suppose, toward a contradiction, that a type–2 firm offers $w_2$ and a
% type–1 firm offers $w_1$ with $w_2 < w_1$.

% Value constancy on each support requires
% \[
% \ell(w_2) M_2(w_2) > \ell(w_1) M_2(w_1), 
% \qquad
% \ell(w_2) Y_1(w_2) < \ell(w_1) Y_1(w_1).
% \]

% Since $M_2(w) = Y_1(w) + \Delta(w) + \eta S(w)$, where
% $\Delta(w)=Y_2(w)-Y_1(w)$ and $S(w)$ is increasing,
% the first inequality becomes
% \[
% \ell(w_2) \bs{Y_1(w_2) + \Delta(w_2) + \eta S(w_2)}
% > \ell(w_1) \bs{Y_1(w_1) + \Delta(w_1) + \eta S(w_1)}
% \]

% Using the second inequality to substitute for $Y_1$, we obtain
% \[
% \ell(w_2) \bs{\Delta(w_2) + \eta S(w_2)}
% > \ell(w_1) \bs{\Delta(w_1) + \eta S(w_1)}
% \]

% Because $\ell(w)$ and $S(w)$ are strictly increasing and
% $w_2<w_1$, and because $y_2(e)\ge y_1(e)$ for all $e$ implies $\Delta(w_2) \le \Delta(w_1)$, the right–hand side weakly exceeds the left–hand side—a contradiction.

% Hence REFs cannot lie strictly below PMFs.  
% Therefore, in any separated-support equilibrium, type–1 must form the lower segment and type–2 the upper segment.

\subsection{Proof of Proposition \ref{prop:BM_disjoint}}
We start by establishing the following Lemmas.

\begin{lemma}\label{lem:overlap}
Suppose there exists an open interval $I\subset\mathbb{R}$ on which both types post with
positive density (i.e., both mix). 
Then there exist constants $\pi_1,\pi_2>0$ such that
\[
\ell(w) M_i(w) = \pi_i \quad \text{for all } w \in I, \;i=1,2,
\]
and, in particular,
\[
\frac{M_2(w)}{M_1(w)} = \frac{\pi_2}{\pi_1}
\quad\text{for all }w\in I,
\]
so the ratio $M_2(w)/M_1(w)$ is constant on $I$.
\end{lemma}

\begin{proof}
Fix a type $i$ that mixes on $I$. 
In a Burdett-Mortensen mixed equilibrium, that type must be indifferent over all wages in its support, so its steady-state value $\pi_i$ does not depend on $w$ within $I$ and satisfies
\[
\pi_i = \ell(w) M_i(w) \quad \text{for all } w \in I
\] 
If both types mix on $I$, then for every $w \in I$ we have
\[
\ell(w) M_1(w) = \pi_1,\qquad
\ell(w) M_2(w) = \pi_2
\]
Dividing the second by the first yields
\[
\frac{M_2(w)}{M_1(w)} = \frac{\pi_2}{\pi_1}
\quad\text{for all }w\in I,
\]
so the ratio $M_2/M_1$ is constant on $I$.
\end{proof}

\begin{lemma}\label{lem:np}
Suppose $\eta>0$ and, on any active interval, $Y_1''(w)=Y_2''(w)\equiv 0$ while $S''(w)\not\equiv 0$.\footnote{For instance, with 
$y_i(e)= y_0 + \xi e$ and $c(e(w),w) = a e^2 - \zeta e w - f(w)$ where $f(\cdot)$ is any function.
} 
Then there is no open interval on which $M_2(w)=c\,M_1(w)$ holds for a constant $c>0$.
\end{lemma}

\begin{proof}
If $Y_2+\eta S=cY_1$ on an open interval, twice differentiating gives $\eta S''=0$, contradicting $\eta>0$ and $S''\not\equiv0$.
\end{proof}

Given Lemmas \ref{lem:overlap}-\ref{lem:np}, the Proposition \ref{prop:BM_disjoint} follows.

\subsection{Proof of Proposition \ref{prop:BM_REFs_lp}}
Under the linearity assumptions stated in Proposition \ref{prop:BM_disjoint}, overlapping supports are impossible; the two supports meet at a single boundary $w_b$. 
We assume separation holds in what follows.

We first consider the case in which PMFs occupy the lower wage segment.

\paragraph{Case 1: PMFs occupy the lower segment}
If PMFs form the lower segment, their value is constant:
\[
\ell(w_b) Y_1(w_b) = \ell(w^r) Y_1(w^r)
\]
Because $F(w^r)=0$ and $F(w_b)=\alpha$,
\[
\ell(w^r) = \frac{\lambda \delta}{(\delta + \lambda)^2}
\qquad
\ell(w_b) = \frac{\lambda \delta}{\bs{\delta + \lambda(1-\alpha)}^2}
\]
and value constancy reduces to
\begin{equation}\label{eq:lower_boundary}
    \frac{Y_1(w_b)}{Y_1(w^r)} = \frac{\ell(w^r)}{\ell(w_b)} = \bp{\frac{\delta + \lambda(1-\alpha)}{\delta + \lambda}}^2 = r_\alpha \in (0,1)
\end{equation}
Under Assumption \ref{ass:BM_decreasing_margin}, $Y_1$ is strictly decreasing on $[w^r, w_b]$, so \eqref{eq:lower_boundary} has a unique solution $w_b$.

Now one must also check that REFs do prefer the upper segment at $w_b$.
A type-2 firm at $w_b$ must not want to deviate down to $w^r$:
\[
\ell(w_b) M_2(w_b) \geq \ell(w^r) M_2(w^r)
\implies 
\frac{M_2(w_b)}{M_2(w^r)} \geq r_\alpha
\]
using $\ell(w^r) / \ell(w_b) = r_\alpha$.
Substituting $M_2(w) = Y_2(w)+\eta S(w)$ yields
\[
\eta S(w_b) \geq r_\alpha Y_2(w^r) - Y_2(w_b)
\]

Since $S(w^r)=0$ and $S(w_b)>0$, we obtain the following threshold for $\eta$, at which REFs switch from being unable to sustain the upper segment (low $\eta$) to preferring it (high $\eta$):
\[
\eta \geq \eta^* = \frac{r_\alpha Y_2(w^r) - Y_2(w_b)}{S(w_b)}
=
\frac{r_\alpha Y_2(w^r) - Y_2(w_b)}{\tilde u(w_b)-b}
\]

\paragraph{Case 2: PMFs occupy the upper segment}
A symmetric analysis can be performed when PMFs occupy the higher segment, and one then obtains the following threshold, below which REFs cannot sustain higher wages than PMFs:
\[
\eta \leq \eta^{**} = \frac{r^*_\alpha Y_2(w^*_b)- Y_2(\bar{w})}{ S(\bar{w}) - r^*_\alpha S(w^*_b)}
\]
where the boundary satisfies $F(w^*_b)=1-\alpha$ (since type--1 firms occupy the upper segment) and
\[
\frac{\ell(w^*_b)}{\ell(\bar{w})} = r^*_\alpha = \bp{\frac{\delta}{\delta+\lambda\alpha}}^2 \in (0,1)
\]
and $\bar{w}$ is the upper boundary of the wage distribution ($F(\bar{w})=1$).

% Finally, note that even if $\eta \geq \eta^*$, a type-2 upper mixed segment exists only if, at the entry wage $w_b$, $Z_{2,\eta} (w_b)>0$ and the margin is locally decreasing:
% \[
% Z'_{2,\eta} (w_b) = Y'_2(w_b) + \eta S'(w_b) < 0
% \]
% which depends on the slope of the productive part $Y_2$ and on how strongly worker surplus responds to the wage.

%%%%%%%%%%%%%%%% DMP %%%%%%%%%%%%%%%%%%%%%%%%
\section{Additional computations and proofs for the DMP model}\label{a:add_computations_DMP}

\subsection{Surplus formulas}\label{app:DMP_surplus}
Start from a generic two-state Bellman system,
\[
A=a+\beta\bs{(1-\delta)A+\delta C},\qquad
C=c+\beta\bs{(1-d)C+dA},
\]
where $d\in(0,1)$ is the transition probability from $C$ to $A$.
Subtracting the two equations yields
\[
A-C=(a-c)+\beta(1-\delta-d)(A-C),
\]
so
\[
A-C=\frac{a-c}{1-\beta(1-\delta-d)}.
\]

\paragraph{Firm surplus}
Applying this formula to $(A,C)=(J,V)$ with $a=y-w$, $c=-\kappa$, and $d=q(\theta)$ gives
\[
S_f=J-V=\frac{y-w+\kappa}{1-\beta\left(1-\delta-q(\theta)\right)}.
\]
We will also use the associated cash-flow (outside-option) term:
\[
(1-\beta)V
=\frac{-\kappa\left(1-\beta(1-\delta)\right)+\beta q(\theta)\left(y-w\right)}
{1-\beta\left(1-\delta-q(\theta)\right)}.
\]

\paragraph{Worker surplus}
For the worker, the relevant transition probability from unemployment is $p(\theta)=\theta q(\theta)$.
For a job paying wage $w$, applying the same formula to $(A,C)=(W,U)$ with $a=w$, $c=b$, and $d=p(\theta)$ yields
\[
S_\omega(w)=W(w)-U=\frac{w-b}{1-\beta\left(1-\delta-\theta q(\theta)\right)}.
\]
When there are two firm types with wages $w$ and $w_{\Rc}$ and shares $(1-\mu,\mu)$, the unemployment value depends only on the average wage offer $\bar w:=(1-\mu)w+\mu w_{\Rc}$ and satisfies
\[
(1-\beta)U=(1-F)b+F\bar w,
\qquad
F:=\frac{\beta \theta q(\theta)}{1-\beta\left(1-\delta-\theta q(\theta)\right)}.
\]

\subsection{Two-type wage system under fixed tightness}\label{app:DMP_twotype_system}
Under the assumptions stated in Section~\ref{sec:DMP} (common $\kappa$, common $q(\theta)$, and common separation probability $\delta$), the bargaining-stage outside values entering the wage equations are affine in wages.
Define
\[
E:=\frac{\beta q(\theta)}{1-\beta\left(1-\delta-q(\theta)\right)},\qquad
F:=\frac{\beta \theta q(\theta)}{1-\beta\left(1-\delta-\theta q(\theta)\right)}.
\]
Using the expression for $(1-\beta)V$ from Appendix~\ref{app:DMP_surplus}, we can write
\[
y-(1-\beta)V=\bp{1-E}y+Ew+\kappa\frac{1-\beta(1-\delta)}{1-\beta\left(1-\delta-q(\theta)\right)}.
\]%
Using the expression for $(1-\beta)U$ from Appendix~\ref{app:DMP_surplus}, we have
\[
(1-\beta)U=(1-F)b+F\bar w,\qquad \bar w=(1-\mu)w+\mu w_{\Rc}.
\]
Substituting these objects into the wage equations \eqref{eq:DMP_rf_regfirm}--\eqref{eq:DMP_rf_respfirm} yields two linear equations in $(w,w_{\Rc})$:
\[
w=\frac{c+d\,w_{\Rc}}{D},
\qquad
w_{\Rc}=\frac{c'+d'\,w}{D'},
\]
where
\[
d=(1-\phi)F\mu,\qquad D=1-\phi E-(1-\phi)F(1-\mu),
\]
\[
d'=(1-\phi_{\Rc})F(1-\mu),\qquad D'=1-\phi_{\Rc}E-(1-\phi_{\Rc})F\mu,
\]
and
\[
c=\phi\bp{(1-E)y+\kappa\frac{1-\beta(1-\delta)}{1-\beta(1-\delta-q(\theta))}}+(1-\phi)(1-F)b,
\]
\[
c'=\phi_{\Rc}\bp{(1-E)y+\kappa\frac{1-\beta(1-\delta)}{1-\beta(1-\delta-q(\theta))}}+(1-\phi_{\Rc})(1-F)b.
\]

\subsection{Proof of Proposition \ref{prop:DMP_rf_wages}}
\begin{proof}
    Appendix~\ref{app:DMP_twotype_system} shows that the wage equations reduce to a linear system in $(w,w_{\Rc})$.
    Solving it yields
    \[
        w = \frac{D'c + dc'}{DD'-dd'} ,\qquad
        w_{\Rc} = \frac{Dc' + d'c}{DD'-dd'}.
    \]
    Moreover, $w_{\Rc} > w$ is equivalent to
    \[
    \frac{D'-d'}{D-d} < \frac{c'}{c}.
    \]
    Since $\phi_{\Rc}>\phi$, we have $c'>c$ if and only if total match surplus is positive at $w=0$, i.e.
    \[
    \frac{y+\kappa}{1-\beta(1-\delta-q(\theta))}>\frac{b}{1-\beta(1-\delta-\theta q(\theta))}.
    \]
    Next, note that
    \[
    D-d=1-\phi E-(1-\phi)F,\qquad D'-d'=1-\phi_{\Rc}E-(1-\phi_{\Rc})F.
    \]
    Since $E,F\in(0,1)$ and $\phi<1$, we have $D-d=1-\phi E-(1-\phi)F>0$ (because $\phi E+(1-\phi)F<\phi+(1-\phi)=1$).
    Because $\theta<1$ implies $E>F$, we have $D'-d'<D-d$ and therefore $(D'-d')/(D-d)<1$.
    To see this, use the definitions
    \[
        E=\frac{\beta q(\theta)}{1-\beta(1-\delta-q(\theta))},
        \qquad
        F=\frac{\beta\theta q(\theta)}{1-\beta(1-\delta-\theta q(\theta))},
    \]
    and compute
    \[
        E>F
        \ \Leftrightarrow\
        \frac{\beta q}{1-\beta(1-\delta-q)}>\frac{\beta\theta q}{1-\beta(1-\delta-\theta q)}
        \ \Leftrightarrow\
        (1-\beta+\beta\delta)(1-\theta)>0.
    \]
    Combining these two inequalities yields $w_{\Rc}>w$.
\end{proof}

\subsection{Proof of Proposition \ref{prop:DMP_w0}}
\begin{proof}
    Let $w_0$ denote the wage when $\mu=0$, i.e., where no responsible firms operate in the market.
    Using the same notation as in previous proofs, we have:
    \[
    w_0 = \frac{c}{D-d}
    \]
    The difference between the wage of regular firms before and after the introduction of responsible firms is:
    \[
    w - w_0 = \frac{c + dw_{\Rc}}{D} - \frac{c}{D-d} = d \frac{c'(D-d) - c(D'-d')}{(DD'-dd')(D-d)} = \bp{w_{\Rc} - w} \frac{d}{D-d}
    \]
    This difference is positive under the same conditions as Proposition \ref{prop:DMP_rf_wages}, since $d>0$ and $D-d>0$.
\end{proof}

\subsection{Proof of Proposition \ref{prop:DMP_wR_monotonicity}}
\begin{proof}
    Recall from the previous proofs the expression:
    \[
    w_{\Rc} - w = \frac{c'(D-d) - c(D'-d')}{DD' -  dd'}
    \]
    The numerator does not depend on $\mu$, so we can focus on $DD' - dd'$.
    This expression can be written as:
    \[
    \bs{1 - \phi E - (1-\phi)(1-\mu) F} \bs{1 - \phi_{\Rc} E - (1-\phi_{\Rc}) \mu F} 
    - (1-\phi)(1-\mu)F (1-\phi_{\Rc})\mu F 
    \]
    where $E = \frac{\beta q(\theta)}{1 - \beta \bp{1 - \delta - q(\theta)}}$ and $F = \frac{\beta \theta q(\theta)}{1 - \beta \bp{1 - \delta - \theta q(\theta)}}$.
    
    This is an affine function in $\mu$, with a coefficient equal to
    \[
    F\Bigl[(1-\phi)\bs{1-\phi_{\Rc}E}-(1-\phi_{\Rc})\bs{1-\phi E}\Bigr]
    \]
    which simplifies to $F(\phi_{\Rc}-\phi)(1-E)>0$.
    
    Thus, the wage gap monotonically decreases with the proportion of firms that have a higher $\phi_{\Rc}$.
\end{proof}

\subsection{Relation to flow welfare}\label{a:flow_welfare}
Define the standard flow welfare measure as
\[
    \mathcal{W} = (1-u)\,y + u\,b - \kappa\,v.
\]
With fixed tightness $\theta$ and fixed masses (e.g.\ under limited entry with fixed $N$ so that $u$ and $v$ are fixed), varying the bargaining weight $\varphi$ changes only the wage $w$, which is a transfer between workers and firms.
Since $y$, $b$, $\kappa$, $u$, and $v$ are all unaffected, $\mathcal{W}$ is invariant to $\varphi$ (and hence to $w$) in this case.
Under free entry, $\varphi$ affects equilibrium tightness $\theta$ and therefore $\mathcal{W}$ through vacancy creation and the resulting changes in $u$ and $v$.
Our stakeholder-value criterion $\bp{N-v}\bp{S_\omega+S_f}$ differs from $\mathcal{W}$ because it evaluates each match relative to the parties' outside states $(U,V)$, which depend on wages and tightness; it is the natural governance object when firms and workers are the relevant stakeholders and outside values matter for bargaining and participation decisions.

\end{document}